\documentclass[linenumbers]{aastex62}

\usepackage{longtable}
\begin{document}
\title{OO Dra: an Algol-type binary formed through an extremely helium-poor mass accretion revealed by asteroseismology}
\correspondingauthor{Xinghao Chen,  Xiaobin Zhang,  and Yan Li}
\email{chenxinghao1003@163.com,  xzhang@bao.ac.cn, and ly@ynao.ac.cn}
\author[0000-0003-3112-1967]{Xinghao Chen}
\affiliation{Yunnan Observatories, Chinese Academy of Sciences, P.O. Box 110, Kunming 650216, China}
\affiliation{Key Laboratory for Structure and Evolution of Celestial Objects, Chinese Academy of Sciences, P.O. Box 110, Kunming 650216, China}
\author[0000-0002-5164-3773]{Xiaobin Zhang}
\affiliation{Key Laboratory of Optical Astronomy, National Astronomical Observatories, Chinese Academy of Sciences, Beijing, 100012, China}
\author{Yan Li}
\affiliation{Yunnan Observatories, Chinese Academy of Sciences, P.O. Box 110, Kunming 650216, China}
\affiliation{Key Laboratory for Structure and Evolution of Celestial Objects, Chinese Academy of Sciences, P.O. Box 110, Kunming 650216, China}
\affiliation{University of Chinese Academy of Sciences, Beijing 100049, China}
\affiliation{Center for Astronomical Mega-Science, Chinese Academy of Sciences, 20A Datun Road, Chaoyang District, Beijing, 100012, China}
\author{Changqing Luo}
\affiliation{Key Laboratory of Optical Astronomy, National Astronomical Observatories, Chinese Academy of Sciences, Beijing, 100012, China}
\author{Xuzhi Li}
\affiliation{Yunnan Observatories, Chinese Academy of Sciences, P.O. Box 110, Kunming 650216, China}
\affiliation{Key Laboratory for Structure and Evolution of Celestial Objects, Chinese Academy of Sciences, P.O. Box 110, Kunming 650216, China}
\author[0000-0001-7566-9436]{Jie Su}
\affiliation{Yunnan Observatories, Chinese Academy of Sciences, P.O. Box 110, Kunming 650216, China}
\affiliation{Key Laboratory for Structure and Evolution of Celestial Objects, Chinese Academy of Sciences, P.O. Box 110, Kunming 650216, China}
\author{Xuefei Chen}
\affiliation{Yunnan Observatories, Chinese Academy of Sciences, P.O. Box 110, Kunming 650216, China}
\affiliation{Key Laboratory for Structure and Evolution of Celestial Objects, Chinese Academy of Sciences, P.O. Box 110, Kunming 650216, China}


\begin{abstract}
Based on the two-minutes TESS data,  we analyzed intrinsic oscillations of the primary component and identified seven confident independent $\delta$ Scuti frequencies ($f_1$,  $f_2$, $f_3$,  $f_4$, $f_7$,  $f_{11}$,  and $f_{12}$).  Both of single-star evolutionary models and mass-accreting models are computed to reproduce the $\delta$ Scuti freqiencies.  Fitting results of them match well with each other.  The stellar parameters of the primary star yielded by asteroseismology are $M$ = $1.92^{+0.10}_{-0.02}$ $M_{\odot}$, $Z$ = 0.011$^{+0.006}_{-0.001}$,  $R$ = $2.068^{+0.050}_{-0.007}$ $R_{\odot}$, $\log g$ = $4.090^{+0.010}_{-0.002}$, $T_{\rm eff}$ = $8346^{+244}_{-320}$ K,  $L$ = $18.65^{+3.31}_{-2.82}$ $L_{\odot}$,  which match well with the dynamic ones by the binary model.  Furthermore,  our asteroseismic results show that OO Dra is another Algol system that has just undergone the rapid mass-transfer stage.  Fitting results of single-star evlutionary models indicate that the pulsator is  helium-poor star with an age of 8.22$^{+0.12}_{-1.33}$ Myr,  and the further mass-accreting models show that the primary star looks like an almost unevolved star formed by an extremely helium-poor mass accretion in Case A evolutionary scenario.
\end{abstract}
\section{Introduction}
Eclipsing binaries with pulsating components are very important objects to understand stellar structure and evolution.  The eclipse nature allow us to derive accurate physical parameters of the components,  and the oscillation frequencies provide significant insight into their interiors as well as opportunities to identify physical processes behind the pulsating nature of the components(Aerts $\&$ Harmanec 2004;  Mkrtichian et al.  2005,  2007).  Various types of pulsating stars have been found in eclipsing binaries,  such as red gaints (Beck et al.  2014; Gaulme et al. 2013,  2014,  2016),  $\gamma$ Dor stars(Maceroni et al. 2013; Guo et al.  2017; Ibanoglu et al.  2018; Guo \& Li 2019), and $\delta$ Scuti stars(Kahraman et al.  2017; Liakos $\&$ Niarchos 2017; Gaulme $\&$ Guzik 2019).  Detailed studies of pulsating stars in eclipsing binaries have been carried out in many literatures (Schmid $\&$ Aerts 2016; Kahraman et al.  2017; Beck et al.  2018;  Zhang et al.  2018; Bowman et al.  2019; Chen et al.  2020). Schmid $\&$ Aerts (2016) presented an asteroseismic study of the $\delta$ Scuti-$\gamma$ Doradus pulsating binary KIC 10080943 and showed that the size of core overshooting and diffusive mixing can be well constrained in the hypothesis of the same age of the two components.  Beck et al.  (2018) analyzed the eccentric binary KIC 9163796 and reported that the two pulsating red giant components are in early and late phases of the first dredge-up event on the red giant branch.  Most of eclipsing binaries with $\delta$ Scuti-type pulsations are found to be Algol-type systems.  The pulsating components exhibit similar pulsation behaviors with those of single $\delta$ Scuti pulsators,  but they have a distinct evolutionary history due to mass transfer between the two components.  Kahraman et al. (2017) presented a comparison between pulsating eclipsing binaries and single $\delta$ Scuti pulsators and found that the $\delta$ Scuti components in binaries pulsate at lower amplitudes and shorter periods than those of single pulsators.  Guo et al.  (2017) analyzed the post mass-transfer binary KIC 9592855 and presented that the core and envelope of the pulsator rotate nearly uniformly and both of their rotation rates are similar to the orbital motion.  Bowman et al.  (2019) investigated the TESS light curve of the oEA system U Gru and showed that tidally perturbed oscillations can occur in p-mode region.  Chen et al.  (2020) carried out an  analysis of the Algol system KIC 10736223 through binary properties and 
asteroseismology,  they reported that KIC 10736223 has just undergone the rapid mass-transfer stage.  Post mass-transfer binaries carry evidence of binary interaction,  thus the study of them offers new and strict constraints to refine stellar evolution theories,  which can help us further to understand the influences of tidal forces and mass transfer processes among the interacting binaries.  However,  mode identification is more difficult for $\delta$ Scuti stars due to the low radial orders.  So far,  comprehensive asteroseismic modeling of $\delta$ Sct-type pusators in post mass-transfer binaries is still lacking. 

OO Dra was first discovered by Biyalieva $\&$ Khruslov (2007) to be an eclipsing binary with an orbital period of 1.23838 days.  The pulsating nature of OO Dra was later found by Dimitrov et al.  (2008) through their BVR passbands light curves.  They made a preliminary frequency analysis of the out-of eclipse photometric data,  and detected a main frequency about 37 cycle days$^{-1}$.  Besides,  Dimitrov et al.  (2008) contributed a spectroscopy of OO Dra,  from which the effective temperature,  surface gravity,  and projected rotational velocity of the primary star are determined to be $T_{\rm eff,1}$ = 8500 K,  $\log g_1$= 4.0 dex,  and $\upsilon_1\sin i$ = $\sim$60 km s$^{-1}$,  respectively.  Zhang et al.  (2014) presented a more comprehensive BV bands photometric observations of OO Dra,  and found that the binary may be a detached system with the less-massive secondary nearly filling its Roche lobe.  They detected two frequencies of 41.87 cycle days$^{-1}$ and 34.75 cycle days$^{-1}$ in both bands of the light residuals,  and confirmed the primary star to be a pulsator with $\delta$ Scuti pulsations. Lee et al.  (2018) presented time-series spectroscopy of OO Dra,  and derived the effective temperature and the projected rotational velocity of the primary component to be 8260 $\pm$ 210 K and 72$\pm$ 5 km s$^{-1}$,  respectively.  Furthermore,  Lee et al.  (2018) modeled their new RV curves and the BV photometric data of Zhang et al.  (2014),  and obtained physical parameters of the primary star to be $M_1$ = 2.031 $\pm$ 0.058 $M_{\odot}$, $R_1$ = 2.078 $\pm$ 0.026 $R_{\odot}$, $\log g_1$ = 4.110 $\pm$ 0.017, $L_1$ = 18.0 $\pm$ 1.9 $L_{\odot}$, and those of the secondary star to be $M_2$ = 0.187 $\pm$ 0.009 $M_{\odot}$, $R_2$ = 1.199 $\pm$ 0.017 $R_{\odot}$, $\log g_2$ = 3.552 $\pm$ 0.026, $L_2$ = 1.99 $\pm$ 0.20 $L_{\odot}$, respectively.  

In this work, we extend works of Zhang et al.  (2014) and Lee et al.  (2018), and perform a more comprehensive asteroseismic analysis for the eclipsing binary OO Dra.  In Section 2, we present pulsational features of the primary component.  We introduce the details of input physics in Section 3,  and elaborate our asteroseismic modelings in Section 4.  Finally,  we conclude and discuss our results in Section 5.
\section{Frequency analysis}
OO Dra was observed in two-minutes cadence mode by TESS satellite during Sectors 20 and 21 from 2019 December 24 to 2020 February 18 with a time span of $\Delta T$ = 58.28 days.  We downloaded the data from the Mikulski Archive for Space Telescopes (MAST,  https://mast.stsci.edu/portal/Mashup/Clients/Mast/Portal.html),  and all TESS data used in this paper can be found in MAST: 10.17909/t9-kfsv-6g77.  From the data file,  we extracted the BJD times and the "PDCSAP$\_$FLUX" fluxes,  which were processed with the Pre-search Data Conditioning Pipeline (Jenkins et al.  2016) to eliminate instrumental trends.  Afterwards,  we removed outliers and normalized the fluxes with the method described in Slawson et al.  (2011).  

Lee et al.  (2018) determined physical parameters of OO Dra by using the B- and V-band photometry light curves,  while observations by TESS are in white light that do not includes colour information.  In this work,  we adopt physical parameters obtained by Lee et al.  (2018).  Using the ephemerides given by Lee et al. (2018),  we computed phases and folded the light curve as illustrated in Figure 1. 

In order to remove the light changes due to eclipses,  and obtain the pulsational light variations,  we adopt a simple approach,  i.e.,  computing a mean curve for the folded the light curve and then subtracting it from the two-minute cadence data.  When computing the mean curve,  the number of bins in phase should not be too large so as to ensure that each bin can contain enough data points.  Given that the 58.28 day's TESS data of OO Dra include 35927 data points,  we use 1500 bins in phase in this work. Moreover, we also use 500 bins and 1000 bins in phase for the folded the light curve,  which will result in slightly different wiggles. However, we find those differences and interpolation algorithms do not impact the extracted frequencies.  In Figure 1,  we show the mean curve with the red line in the upper panel,  and show the light residuals in plots of magnitude versus phase in the lower panel.  Afterwards,  we perform a multiple frequency analysis of the light residuals with the software Period04 (Lenz $\&$ Breger 2005) to investigate the pulsation nature in details.  No peaks higer than 100 day$^{-1}$ are found,  thus we perform further frequency extractions in the range of 0$-$100 day$^{-1}$.

At each step of the iteration,  we pick out the frequency with the highest amplitude and carry out a multi-period least-square fit to the data using all frequencies that have been already detected.  The data are then pre-whitened and the residuals were used for further analysis.  Following Breger et al.  (1993), we adopt the empirical threshold of the signal-to-noise ratio S/N = 4 for a frequency to be accepted as significant.  Finally,  we extracted a total of 45 frequencies with S/N $>$ 4,  and listed them in Table 1.  The uncertainties of frequencies and amplitudes are calculated using Monte Carlo simulations as described in Fu et al. (2013),  and noises are computed in the range of 2 day$^{-1}$ around each frequency.  Figure 2 depicts Fourier amplitude spectra of the light residuals,  with the original spectrum illustrated in the upper panel and the residual spectrum of the 45 frequencies after pre-whitening illustrated in the lower panel.

Following P{\'a}pics et al.  (2012) and Kurtz et al.  (2015),  we examine the extracted frequencies to search for the possible orbital harmonics and liner combinations in the form of $f_i = f_j \pm mf_{\rm orb}$ or $f_i = mf_j + nf_k$,  where $f_j$ and $f_k$ are the parent frequencies,  $f_i$ is the combination term,  $m$ and $n$ are integers,  and $f_{\rm orb}$ = 0.8075 day$^{-1}$.  We accepted a peak as a combination if the amplitudes of both parent frequencies are larger than that of the presumed combination term,  and the difference between the observed frequency and the predicted frequency is smaller than the frequency resolution 1.5/$\Delta$T= 0.027 day$^{-1}$ (Loumos $\&$ Deeming 1978).  We identify a total of 33 such frequencies,  and mark them in Table 1.  Besides,  we accepted the lower amplitude frequency as an unresolved peak if two frequencies are closer than 1.5/$\Delta$T= 0.027 day$^{-1}$.  Finally,  seven confident independent frequencies $f_1$,  $f_2$,  $f_3$, $f_4$,  $f_7$,  $f_{11}$,  and $f_{12}$ are retained as shown in Table 1.  

\section{Input physics}
Our models are computed with the one-dimensional stellar evolution code Modules for Experiments in Stellar Astrophysics (MESA, version 10398,  Paxton et al.  2011,  2013,  2015,  2018).  In particular,  we use its submodule "pulse$\_$adipls" to generate evolutionary models of stars and calculate the corresponding adiabatic frequencies (Christensen-Dalsgaard 2008).

In the calculations,  we choose the 2005 update of the OPAL equation of state tables (Rogers \& Nayfonov 2002).  We adopt the OPAL opacity tables of Iglesias \& Rogers (1996) in high temperatures region,  and use tables of Ferguson et al. (2005) in the low temperature region.  The solar metal composition AGSS09 (Asplund et al. 2009) is used as the initial ingredient in metal,  and a simple photosphere is adopted for the atmospheric surface boundary conditions.  The classical mixing length theory of B$\ddot{\rm o}$hm-Vitense (1958) with the solar value of $\alpha$ = 1.9 (Paxton et al. 2013) is used in the convective region.  In addition,  we do not consider effects of the stellar rotation,  the element diffusion,  and the convective overshooting,  as well as magnetic fields on the structure and evolution of the star in our models (for more details,  referring to the Appendix).
\section{Asteroseismic analysis}
\subsection{Single-star Evolutiuonary Models}
Our single-star evolutionary grid considers variations of the stellar mass $M$,  the helium abundance $Y$,  and the metallicity $Z$,  which dictate the evolutionary track of a star.  Therein,  the stellar mass $M$ varies from 1.80 $M_{\odot}$ to 2.20 $M_{\odot}$ in an interval of 0.02 $M_{\odot}$,  and metallicities $Z$ varies from 0.005 to 0.030 in an interval of 0.001.  For the helium abundance $Y$,  we adopt three different values: the extremely helium-poor abundance 0.220,  the moderate helium abundance $0.249+1.33Z$ (Li et al.  2018) as a function of $Z$,  and the helium-rich abundance 0.300.   

Each star in the grid is computed evolving from pre-main sequence stage to the evolutionary status where the central hydrogen of the star exhausts ($X_c<$ 1$\times$10$^{-5}$).  Based on the results of binary model given by Lee et al. (2018),  we use 8000 K $< T_{\rm eff} <$ 8600 K and 1.95 $R_{\odot}$ $<$ $R$ $<$ 2.15 $R_{\odot}$ as the observational constraints.  When a star evolves along its evolutionary track,  we calculate adiabatic frequencies of the radial oscillations ($\ell = 0$) and nonradial oscillations with $\ell$ = 1 and 2 for all stellar models that meet with the observational constraint.  Beacuse of the cancellation effects of the surface geometry,  oscillation modes with higher degree are hardly visible,  thus oscillation modes with $\ell \ge 3$ are not included in this work.

In general,  the component stars in binaries rotate along with the orbital motion.  Following Chen et al. (2020),  we consider the rotation period $P_{\rm rot}$ of the star as the fourth adjustable parameter,  varying from 1.0 days to 1.50 days with a step of 0.01 days.  According to the theory of stellar oscillations,  rotation will result in that each nonradial oscillation mode with $\ell$ splits into $2\ell+1$ different frequencies.  The general expression of the first-order effect of rotation on pulsation was deduced to be
\begin{equation}
\nu_{\ell,n,m}=\nu_{\ell,n}+m\delta\nu_{\ell,n}=\nu_{\ell,n}+\beta_{\ell,n}\frac{m}{P_{\rm rot}}
\end{equation}
(Aerts et al. 2010).  In the equation,  $\ell$,  $n$,  and $m$ are three indices that characterize oscillation modes,  and $\delta\nu_{\ell,n}$ is the splitting frequency.  According to equation (1),  the effect of the rotation on pulsation can be completely dictated by the constant $\beta_{\ell,n}$.  Aerts et al.  (2010) derived the general expression of $\beta_{\ell,n}$ to be
\begin{equation}
\beta_{\ell, n}=\frac{\int_{0}^{R}(\xi_{r}^{2}+L^{2}\xi_{h}^{2}-2\xi_{r}\xi_{h}-\xi_{h}^{2})r^{2}\rho dr}
{\int_{0}^{R}(\xi_{r}^{2}+L^{2}\xi_{h}^{2})r^{2}\rho dr},
\end{equation}
where $\xi_r$ is the radial displacement,  $\xi_h$ is the horizontal displacement,  $\rho$ is the local density of the star,  and $L^2= \ell(\ell+1)$.  According to the above analyses,  each dipole mode splits into three different frequencies with $m$ = -1, 0, and +1, respectively,  and each quadrupole mode splits into five different frequencies with $m$ = -2,  -1,  0,  +1,  and +2,  respectively.

Then we compare frequencies between models and observations according to 
\begin{equation}
S^{2}=\frac{1}{k}\sum(|\nu_{\rm mod,i}-\nu_{\rm obs, i}|^{2}),
\end{equation}
in which $\nu_{\rm mod,i}$ and $\nu_{\rm obs,i}$ represent a pair of matched model-observed frequencies,  and $k$ is the number of the observed frequencies.  Because of no preconceived idea of identifications of the $\delta$ Scuti frequencies,  we adopt the random fitting algorithm,  and consider the model frequency nearest to observations as the most probably matched model counterpart. 

Figures 3 and 4 show plots of fitting results $S_{\rm m}^{2}$ versus various physical parameters.  In the figures,  circles in red,  blue,  and black are used to mark theoretical models with $Y$ = 0.220,  0.249+1.33$Z$,  and  0.300,  respecitively.  Similar to Chen et al. (2016),  we find that the solution concentrates on a small parameter space for a given evolutionary track. Thus we pick out the one model with the minimum value of $S^2$ along the evolutionary track,  and denote the minimum value with the symbol $S_{\rm m}^2$.  We use horizontal lines to mark the position of $S^2_{\rm m}$ = 0.10,  which corresponds to the square of the frequency resolution $1.5/ \Delta T$.  

Figures 3(a) depicts the plot of $S_{\rm m}^{2}$ versus the stellar mass $M$.  It can be clearly seen in the figure that theoretical models with poorer helium abundance has higher mass,  i.e.,  1.80$-$1.86 $M_{\odot}$ for $Y$ = 0.300,  1.80$-$1.92 $M_{\odot}$ for $0.249+1.33Z$,  and 1.90$-$2.04 $M_{\odot}$ for $Y$ = 0.220,  respectively.  Lee et al.  (2018) determined the mass of the primary star to be 2.03 $\pm$ 0.06 $M_{\odot}$.  We therefore consider the 61 models with $Y=0.220$ above the horizontal  line as our preferred models,  and list them in Table 2.  Among these models,  Model A44 has the minimum value of $S_{\rm m}^2$,  we then pick out it as the optimal single-star evolutionary model in this work.

Figures 3(b) and 3(c) depict plots of $S_{\rm m}^2$ versus the metallicity $Z$ and the rotational period $P_{\rm rot}$,  respectively.  In the figures,  we noticed that values of $Z$ and $P_{\rm rot}$ show excellent convergence.  Values of $Z$ converge well to 0.014$^{+0.001}_{-0.003}$,  and those of $P_{\rm rot}$ converge well to 1.17$^{+0.01}_{-0.02}$ days.  

Figure 3(d) depicts the plot of $S_{\rm m}^{2}$ as a function of the age of the star.  As illustrated in the figure,  ages of all preferred models converge well to $8.22^{+0.12}_{-1.33}$ Myr.  Similar to KIC 10736223 (Chen et al.  2020),  the primary component of OO Dra also looks like an almost unevolved star near zero-age main sequence.  We therefore identify OO Dra to be another binary system that has just past a rapid mass-transfer stage.

Figures 4(a)-(d) present plots of $S_{\rm m}^2$ as a function of various global stellar parameters,  $\log g$,  $R$,  $T_{\rm eff}$,  and $L$,  respectively.  Values of $\log g$ and $R$ converge  to 4.092$^{+0.007}_{-0.004}$ and 2.075$^{+0.035}_{-0.014}$ $R_{\odot}$,  respectively.  The convergence of $T_{\rm eff}$ and $L$ are relatively worse,  i.e.,  $T_{\rm eff}$ = 8065$^{+528}_{-45}$ K and $L$ = 16.37$^{+5.17}_{-0.47}$ $L_{\odot}$.

Finally,  fundamental physical parameters of the primary star yielded by asteroseismic fittings of single stars are listed in the second column of Table 3.  Table 4 lists model frequencies of the best-fitting single-star evolutionary model,  and Table 5 shows comparisons between the observed frequencies and their corresponding model counterparts.  Based on the comparisons,  $f_{2}$, $f_{7}$ and $f_{11}$ are identified as three dipole modes,  and $f_{1}$,  $f_{3}$, $f_{4}$, and $f_{12}$ as four quadrupole modes.  Furthermore,  we find that $f_7$ and $f_2$ are $m$ = -1 and 0 components of one triplet,  and $f_{12}$ and $f_1$ are $m$ = +1 and +2 components of one quintuplet.
\subsection{Mass-accreting Models}
In Section 4.1,  we analyzed the fitting results of single-star evolutionary models and found that OO Dra might be another binary system that has just undergone a rapid mass-transfer stage.  Given that there are many uncertainties in binary evolution models,  especially for the mass-transfer process,  we model the mass-transfer process in a simlar approach with the work of Chen et al.  (2020).  We evolve a single star to the position where the rapid mass-transfer process may occur ( Case A or Case B binary evolution, Han et al.  2000; Chen et al.  2017),  and then artificially accrete mass at a rate of 10$^{-6}$ $M_{\odot}$yr$^{-1}$ until masses of the accretor up to the given values.   We considered the initial mass $M_1$ of the accretor between 0.40 $M_{\odot}$ and 1.00 $M_{\odot}$ in an interval of 0.1 $M_{\odot}$,  and the final mass $M_2$ of the accretor between 1.90 $M_{\odot}$ and 2.20 $M_{\odot}$ in an interval of 0.02 $M_{\odot}$.  For each pair of them,  we adopt two schemes of the mass accretion,  i.e.,  starting at the age of 0.5 Gyr (the donor on the main sequence stage,  Case A binary evolution) and 1.0 Gyr (the central hydrogen of the donor exhaustion,  Case B binary evolution).  Moreover,  the fitting results of single-star evolutionary models indicate that the primary component of OO Dra is a helium-poor star.  The primordial cosmological helium abundance is determined by Planck Collaboration et al.  (2016) to be 0.249.  Therefore,  we choose the moderate helium abundace 0.249+1.33$Z$ as the base helium abundance of the accretor,  and adopt a helium-poor mass accretion,  i.e.,  accreting 0.85 times of the base helium abundance to the surface of the accretor.  In addition,  we consider $Z$ between 0.005 and 0.030 in an interval of 0.001,  and $P_{\rm rot}$ between 1.00 days and 1.50 days in an interval of 0.01 days.

We compare frequencies between model and observations accroding to equation (3),  and show changes of $S_{\rm m}^2$ as a function of various physical parameters of the mass-accreting models in Figures 5 and 6.  In  the figures,  circles in red and blue correspond to accreting models of Case A and Case B binary evolutionary scenario,  respecitively.  It can be noticed in the figures that the Case A evolutionary scenario reproduces the $\delta$ Scuti frequencies better than the Case B evolutionary scenario.  The circles above the horizontal lines correspond to 85 preferred models of Case A evolutionary scenario in Table 6.  

Figures 5(a)-(d) present changes of fitting results $S_{\rm m}^2$ as a function of $Z$,  $M_1$,  $M_2$,  and $P_{\rm rot}$,  respectively.  It can be clearly seen in the figures that values of them converge well to 0.011$^{+0.006}_{-0.001}$,  0.70$\pm$ 0.1 $M_{\odot}$,  1.92$^{+0.10}_{-0.02}$ $M_{\odot}$,  and 1.16$^{+0.02}_{-0.01}$ days, respectively.

Figures 6(a)-(d) depict changes of $S_{\rm m}^2$ as a function of global physical parameters $\log g$,  $R$,  $T_{\rm eff}$,  and $L$,  respectively.  Therein,  $\log g$ and $R$ exhibit good convergence to 4.090$^{+0.010}_{-0.002}$ and 2.068$^{+0.050}_{-0.007}$ $M_{\odot}$,  respectively.  While $T_{\rm eff}$ and $L$ cover a wide range.  Values of $T_{\rm eff}$ vary from 8026 K to 8590 K and those of $L$ vary from 15.83 $L_{\odot}$ to 21.96 $L_{\odot}$.  We listed the parameters in third column of Table 3,  and find them meeting well with those of the single-star evolutionary models.

Figure 7 presents changes of $S_{\rm m}^2$ as a function of tge of mass-accreting models,  where tge is the evolutionary time since the rapid mass-accretion ends.  In the figure,  tge of the preferred models converge well to $0.49^{+0.18}_{-0.35}$ Myr,  further confirms that OO Dra is an Algol system that has just past the rapid mass-transfer stage. 

Finally,  we list comparisons between model frequencies of the optimal mass-accreting model and observations in Table 7.  In the table,  we notice that identifications of the $\delta$ Scuti frequencies are the same with those of single-star evolutionary scenario. 

\section{\rm Summary and Discussions}
In this work,  we presented simultaneous frequency analyses and asteroseismic modelings for the Algol-type eclipsing binary OO Dra.  We performed a multiple frequency analysis of the eclipse-subtracted TESS light residuals,  and obtained seven confident independent frequencies ($f_1$,  $f_2$,  $f_3$,  $f_4$, $f_7$, $f_{11}$,  and $f_{12}$).  Therein,  $f_1$ and $f_3$ have been detected by Zhang et al. (2014) from B- and V-band pulsational light curves,  while $f_2$,  $f_4$,  $f_7$,  $f_{11}$,  and $f_{12}$ are five new frequencies not detected before.  These frequencies range from 32.4538 cycle days$^{-1}$ to 41.8669 cycle days$^{-1}$,  the $\delta$ Sct-type pulsational nature of OO Dra is further confirmed.

Two grids of theoretical models,  the single-star evolutionary model grid and the mass-accreting model grid,  are computed to reproduce the seven $\delta$ Scuti frequencies.  Due to no preconceived idea of mode identifications of the $\delta$ Scuti frequencies,  a random fitting algorithm is adopted in this work.  Fitting results of mass-accreting models are found to agree well with those of single-star evolutionary models.  Fundamental physical parameters of the primary star yielded by asteroseismology are $M$ = $1.92^{+0.10}_{-0.02}$ $M_{\odot}$, $Z$ = 0.011$^{+0.006}_{-0.001}$,  $R$ = $2.068^{+0.050}_{-0.007}$ $R_{\odot}$, $\log g$ = $4.090^{+0.010}_{-0.002}$, $T_{\rm eff}$ = $8346^{+244}_{-320}$ K,  $L$ = $18.65^{+3.31}_{-2.82}$ $L_{\odot}$,  which match well the results of binary model given by Lee et al.  (2018). 

The primary component of OO Dra is found to be almost an unevolved star near the zero-age main sequence.  Ages of the primary star are determined to be $8.22^{+0.12}_{-1.33}$ Myr for single-star evolutionary models.  Values of tge are determined to be $0.49^{+0.18}_{-0.35}$ for mass-accreting models.  Considering that OO Dra is a detached system with the cool secondary almost filling its Roche lobe (Zhang et al. 2014; Lee et al. 2018),  OO Dra may be another Algol system that has just undergone a rapid mass-transfer stage.

Besides,  our asteroseismic results show that OO Dra is an Algol-type eclipsing binary formed through a helium-poor mass accretion in Case A binary evolutionary scenario.  Figures 8(a) and 8(b) show profiles of Brunt$-$V$\ddot{\rm a}$is$\ddot{\rm a}$l$\ddot{\rm a}$ frequency $N$, characteristic acoustic frequencies $S_{\ell}$ ($\ell$ = 1 and 2),  and hydrogen abundance $X_{\rm H}$ in the best-fitting single-star model (Model A44) and mass-accreting model (Model B21),  respectively.  As shown in the figures,  the dominant difference between the two models is the three bumps in $N$ inside the mass-accreting model.  In particular,  we find that bumps 1 and 2 result from the evolution of the star,  while the highest bump3 is caused by the elemental abundance gradient left behind during the process of helium-poor mass accretion.  The seven $\delta$ Scuti frequencies of OO Dra are found to be p modes,  which mainly propagate in the envelope of the star.  However for the three bumps,  they are not located in the p-mode propagation zones  where $\omega^2 > N^2$ and $\omega^2 > S_{\ell}^2$.  No frequency can reach to the bumps,  that is to say,  the $\delta$ Scuti frequencies are not qualified to distinguish those bumps.  Furthermore,  we introduce an important asteroseismic parameter, the acoustic radius $\tau_0$,  which is the sound travel time between the surface and the center of the star.  The acoustic radius $\tau_0$ is usually used to characterize properties of the stellar envelope (Ballot et al.  2004; Miglio et al.  2010; Chen et al.  2016),  and defined by Aerts et al.  (2010) as 
\begin{equation}
\tau_0=\int_0^R\frac{dr}{c_s}.
\end{equation}
In the equation,  $c_{\rm s}$ is the adiabatic sound speed and $R$ is the stellar radius.  The acoustic radius $\tau_0$ yielded by asteroseismic fittings of mass-accreting models is 2.315$^{+0.009}_{-0.013}$ hr,  which agree well with the value 2.304$^{+0.020}_{-0.003}$ hr of single-star evolutionary models.  In the Hertzsprung–Russell diagram, the instability strip of $\delta$ Scuti stars largely overlaps with that of $\gamma$ Dor stars (Balona 2011; Henry et al. 2011; Uytterhoeven et al. 2011; Xiong et al. 2016).  The $\gamma$ Dor-type pulsations allow us to probe the interiors of the star,  such as chemical mixing (Miglio et al. 2008) and rotation (Van Reeth et al. 2015,  2016,  2018).  The $\delta$ Sct-$\gamma$ Dor hybrid behavior is found to be common for A- and F-type stars (Grigahcène et al. 2010; Balona et al. 2015).  A number of eclipsing binaires with hybrid pulsators have been identified and analyzed,  such as  KIC 9592855 (Guo et al.  2017),  KIC 7385478 (Guo \& Li 2019),  KIC 8113154 (Zhang et al.  2019),  KIC 9850387 (Zhang et al.  2020). Due to the presence of both p and g modes,  eclipsing binaries with hybrid $\delta$ Sct-$\gamma$ Dor pulsators will be very promising objects to understand the influences of mass transfer processes.

Finally,  the rotaion period $P_{\rm rot}$ of the primary star is determined to be $1.17^{+0.01}_{-0.02}$ days for single-star evolutionary models and $1.16^{+0.02}_{-0.01}$ days for mass-accreting models.  The spin of the primary star is slightly faster than the synchronous value.  Saio (1981) and Dziembowski \& Goode (1992) derived that the first-order effect of rotation on pulsation is in proportion to $1/P_{\rm rot}$,  and the second-order effect on pulsation is in proportion to $1/(P_{\rm rot}^2\nu_{\ell,n})$.  The ratio is deduced to be in the order of $1/(P_{\rm rot}\nu_{\ell,n})$.  For OO Dra,  $\nu_{\ell,n}$ varies from 32.4538 day$^{-1}$ to 41.8669 day$^{-1}$.  The second-order effect of rotation on pulsation is much less than that of the first-order one.  Therefore,  the second-order effect of rotation on pulsation is not ncluded in this work.

\acknowledgments
We are sincerely grateful to the anonymous referee for instructive advice and productive suggestions. This paper includes data collected with the TESS mission, obtained from the MAST data archive at the Space Telescope Science Institute (STScI).  Funding for the TESS mission is provided by the NASA Explorer Program.  The authors sincerely acknowledge them for providing such excellent data.  This work is supported by the B-type Strategic Priority Program No. XDB41000000 funded by the Chinese Academy of Sciences.  The authors acknowledge supports of the National Natural Science Foundation of China (grant NO. 11803082 to Chen X.H.,  11833002 and 11973053 to Zhang X.B.,  11333006 and 11521303 to Li Y.,  11803050 to Luo C.Q.,  and 11833006 to Su J.).  We acknowledge the science research grants from the China Manned Space Project with NO. CMS-CSST-2021-B06 and CMS-CSST-2021-B07.  Chen X.H.  also gratefully acknowledges the supports of the Yunnan Fundamental Research Projects and the West Light Foundation of The Chinese Academy of Sciences.  Zhang X.B.  acknowledges the support of Sichuan Science and Technology Program (2020YFSY003). They appreciate the computing time granted by the Yunnan Observatories, and provided on the facilities at the Yunnan Observatories Supercomputing Platform.  They also gratefully acknowledge the "PHOENIX Supercomputing Platform" jointly operated by the Binary Population Synthesis Group and The Stellar Astrophysics Group at Yunnan Observatories, Chinese academy of Sciences.

\appendix 
\section{Inlist files used in this work (Version 10398)}
\subsection{\rm The inlsit file for single-star evolution }
\begin{verbatim}
!inlist1
&star_job
 astero_just_call_my_extras_check_model = .true.
 show_log_description_at_start = .false.
 create_pre_main_sequence_model = .true.
 change_lnPgas_flag = .true.
 new_lnPgas_flag = .true.
 change_initial_net = .true.
 new_net_name = 'o18_and_ne22.net'
 kappa_file_prefix = 'a09'
 kappa_lowT_prefix = 'lowT_fa05_a09p'
 initial_zfracs = 6
/!end of star_job namelist
&controls
 initial_mass = 1.94
 initial_z = 0.014
 initial_y = 0.220
 
 MLT_option = 'ML1'  
 mixing_length_alpha = 1.90
 calculate_Brunt_N2 = .true.
 use_brunt_gradmuX_form = .true.
 which_atm_option = 'simple_photosphere' !default 

 max_number_backups = 50
 max_number_retries = 100
 max_model_number = 80000
 history_interval = 1
 xa_central_lower_limit_species(1) = 'h1'
 xa_central_lower_limit(1) = 1d-5
 max_num_profile_models = 80000

 use_other_mesh_functions = .true.
 mesh_delta_coeff = 0.90
 M_function_weight = 50
 max_center_cell_dq = 1d-10
 Lnuc_div_L_zams_limit = 0.95d0
 max_allowed_nz = 80000
 varcontrol_target = 1d-5
 max_years_for_timestep = 1d6 !for main sequence models (1d3 for pre-main sequence models)
/!end of controls namelist
\end{verbatim}

\subsection{\rm The inlsit file for mass accretion from $M_1$ to $M_2$}
\begin{verbatim}
!inlist2
&star_job
 show_log_description_at_start = .false.
 create_pre_main_sequence_model = .true.
 save_model_when_terminate = .true.
 save_model_filename = 'final_mass.mod'
 change_lnPgas_flag = .true.
 new_lnPgas_flag = .true.
 change_initial_net = .true.
 new_net_name = 'o18_and_ne22.net'
 kappa_file_prefix = 'a09'
 kappa_lowT_prefix = 'lowT_fa05_a09p'
 initial_zfracs = 6
/!end of star_job namelist
&controls
 initial_mass = 0.70
 initial_z = 0.011
 initial_y = 0.26363

 MLT_option = 'ML1'  
 mixing_length_alpha =    1.90
 calculate_Brunt_N2 =.true.
 use_brunt_gradmuX_form = .true.
 radiation_turbulence_coeff = 0 
 which_atm_option = 'simple_photosphere' !default 

 max_number_backups = 50
 max_number_retries = 100
 max_model_number = 80000
 history_interval = 1
 max_num_profile_models = 80000

 use_other_mesh_functions= .true.
 mesh_delta_coeff = 0.90
 M_function_weight = 50
 max_center_cell_dq = 1d-10
 max_allowed_nz = 80000
 varcontrol_target = 1d-5
 max_years_for_timestep = 1d6 !5d3 if star_age > 5d8

 mass_change = 1d-6 !if star_age > 5d8
 accrete_same_as_surface = .false. !if star_age > 5d8
 accrete_given_mass_fractions = .false. !if star_age > 5d8
 accretion_h1 = 0.7649145 !if star_age > 5d8
 accretion_h2 = 0d0 !if star_age > 5d8
 accretion_he3 = 0d0 !if star_age > 5d8
 accretion_he4 = 0.2240855 !if star_age > 5d8
 accretion_zfracs = 6 !if star_age > 5d8
 star_mass_max_limit = 1.92
/!end of controls namelist
\end{verbatim}

\subsection{\rm The inlsit file for evolution of the accreted models }
\begin{verbatim}
!inlist3
&star_job
 astero_just_call_my_extras_check_model = .true.
 show_log_description_at_start = .false.
 load_saved_model = .true.
 saved_model_name = 'final_mass.mod'
 new_lnPgas_flag = .true.
 change_initial_net = .true.
 new_net_name = 'o18_and_ne22.net'
 kappa_file_prefix = 'a09'
 kappa_lowT_prefix = 'lowT_fa05_a09p'
 initial_zfracs = 6
 set_initial_age = .true.
 initial_age = 0 
/!end of star_job namelist
&controls
 initial_mass = 1.92
 initial_z = 0.011
 initial_y = 0.26363

 MLT_option = 'ML1'  
 mixing_length_alpha = 1.90
 calculate_Brunt_N2 =.true.
 use_brunt_gradmuX_form = .true.
 which_atm_option = 'simple_photosphere' !default

 max_number_backups = 50
 max_number_retries = 100
 max_model_number = 80000
 history_interval = 1
 max_num_profile_models = 80000
 xa_central_lower_limit_species(1) = 'h1'
 xa_central_lower_limit(1) = 1d-5

 use_other_mesh_functions= .true.
 mesh_delta_coeff = 0.90
 M_function_weight = 50
 max_allowed_nz = 80000
 max_center_cell_dq = 1d-10
 varcontrol_target = 1d-5
 max_years_for_timestep = 1d6 !for main sequence models (1d3 for pre-main sequence models)
/!end of controls namelist
\end{verbatim}

\begin{figure*}
\centering
\includegraphics[width=0.9\textwidth, angle = 0]{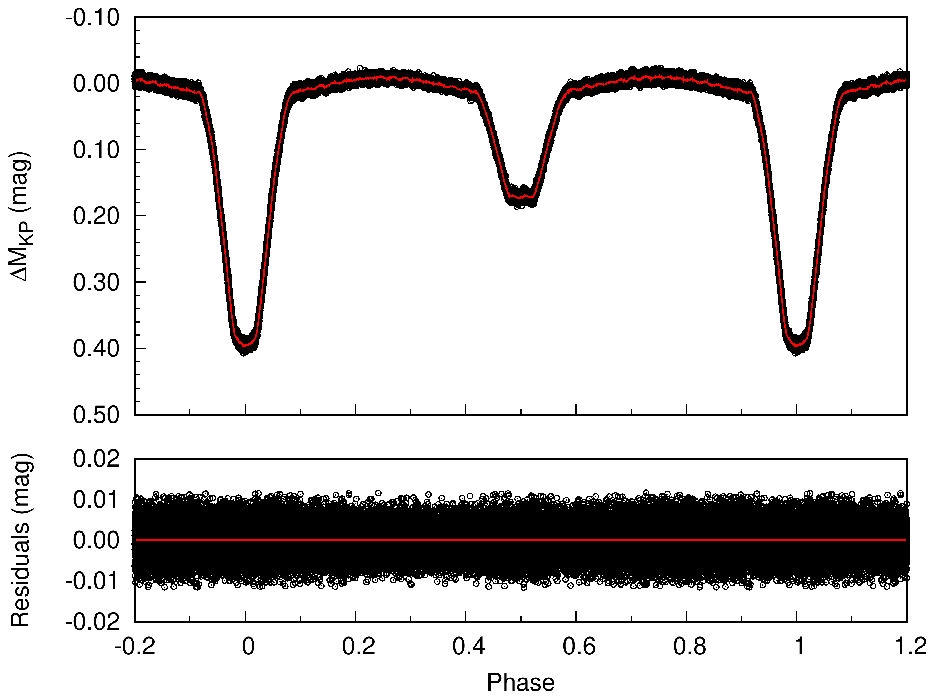}
  \caption{\label{Figure 1}  Light curve and O-C residuals of OO Dra.  The upper panel presents the light curve (black circles),  and the lower panel presents the O-C residuals.  The red line in the upper panel denotes the synthesis mean curve.}
\end{figure*}
\begin{figure*}
\centering
\includegraphics[width=0.9\textwidth, angle = 0]{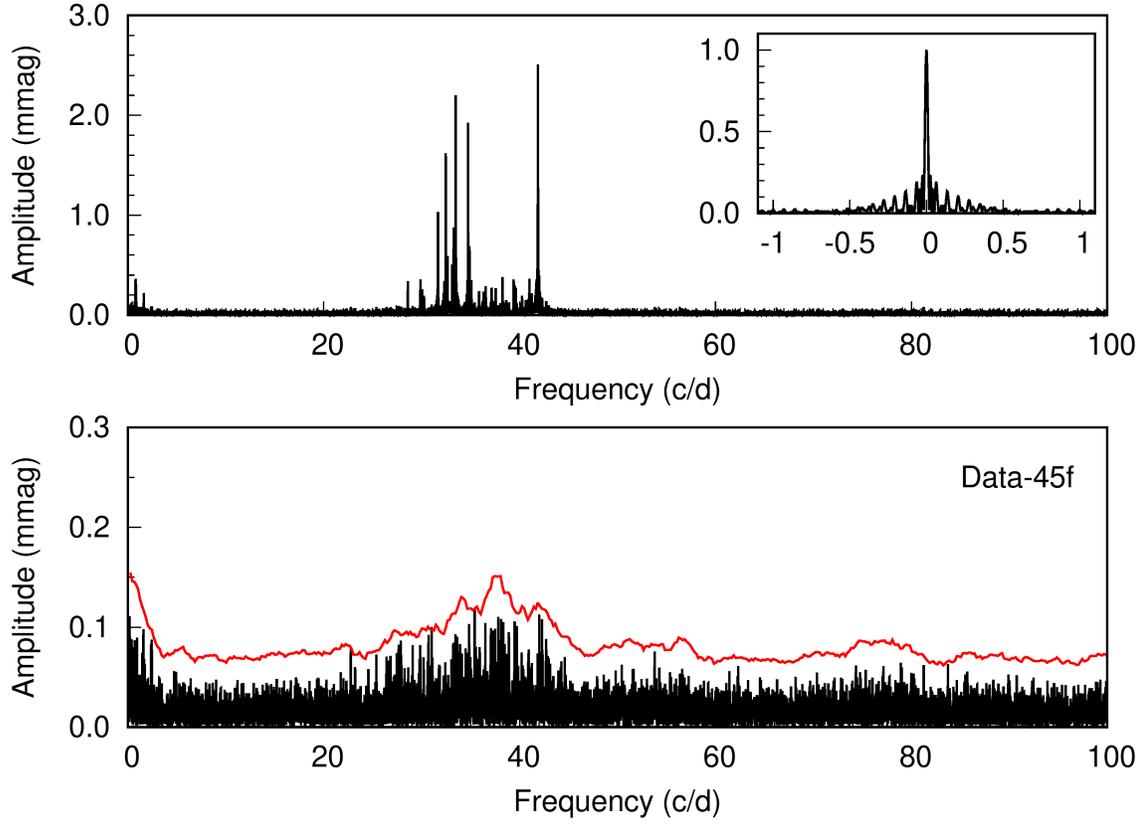}
  \caption{\label{Figure 2}  Fourier amplitude spectrum of the light residuals for OO Dra. The upper panel shows the original spectrum. The inset panel in the upper panel presents the window spectrum.  The lower panel shows the residual spectrum after the 45 detected frequencies were subtracted.  The red line denotes the level of S/N = 4.}
\end{figure*}
\begin{figure*}
\centering
\includegraphics[width=0.9\textwidth, angle = 0]{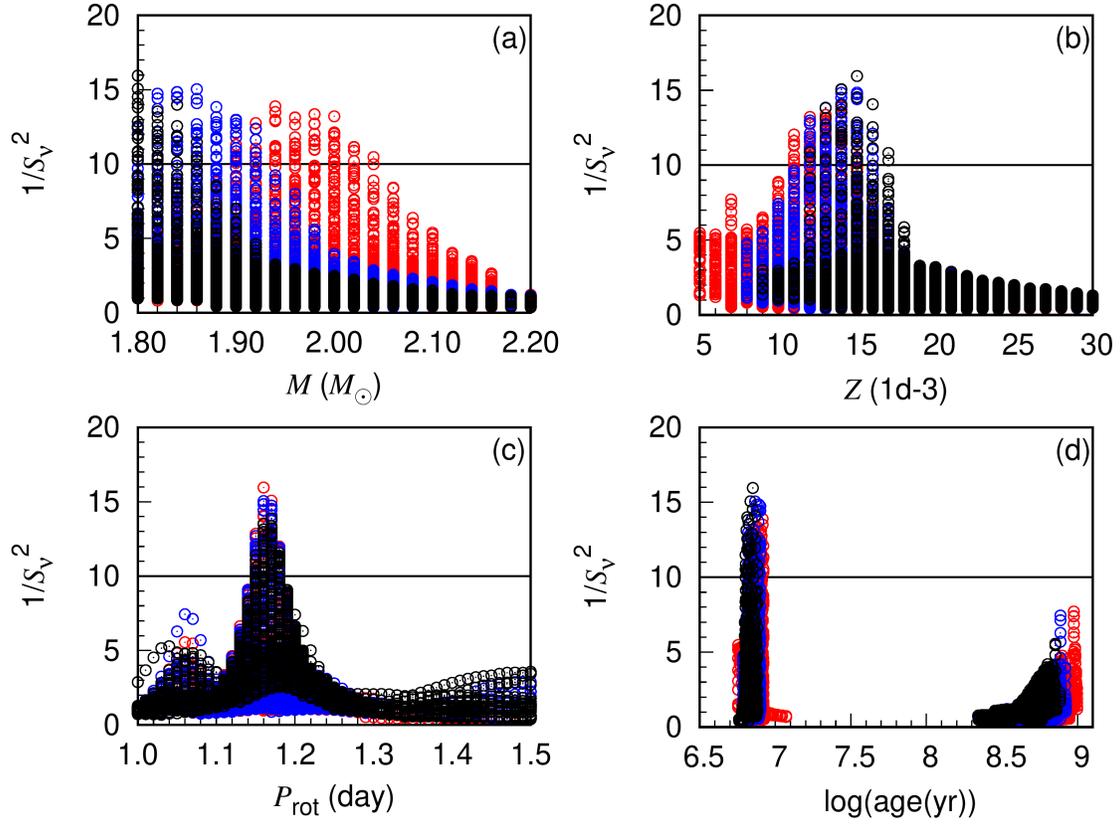}
  \caption{\label{Figure 3} Plots of fitting results $S_{\rm m}^2$ as a function of $M$,  $Z$, $P_{\rm rot}$,  and the age of single-star evolutionary models,  respectively.  The circles in red,  blue  and black correspond to theoretical models with $Y$ = 0.220,  0.249+1.33$Z$,  and  0.300,  respecitively.  The horizontal lines show the position of $S_{\rm m}^2$ = 0.10.}
\end{figure*}
\begin{figure*}
\centering
\includegraphics[width=0.9\textwidth, angle = 0]{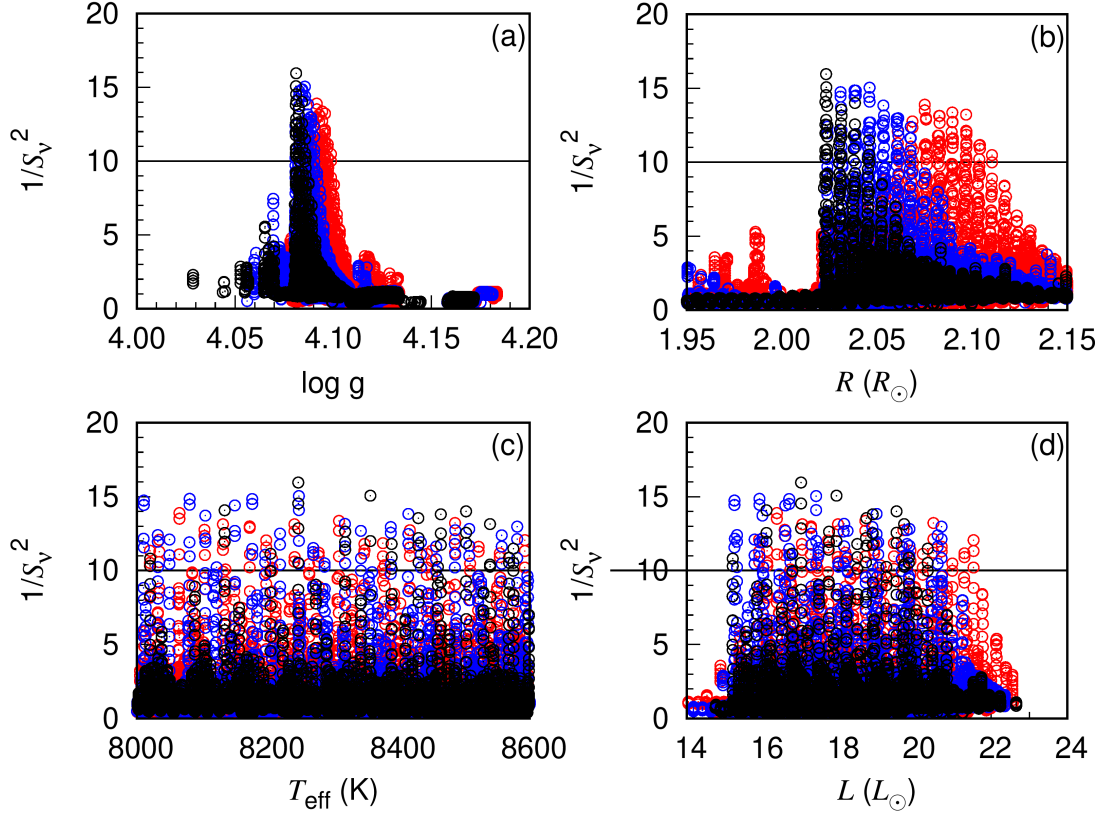}
  \caption{\label{Figure 4} Plots of fitting results $S_{\rm m}^2$ versus global parameters of single-star evolutionary models: $\log g$,  $R$,  $T_{\rm eff}$,  and $L$,  respectively.  The circles in red,  blue,  and black correspond to theoretical models with $Y$ = 0.220,  0.249+1.33$Z$,  and  0.300,  respecitively.  The horizontal lines show the position of $S_{\rm m}^2$ = 0.10.}
\end{figure*}
\begin{figure*}
\centering
\includegraphics[width=0.9\textwidth, angle = 0]{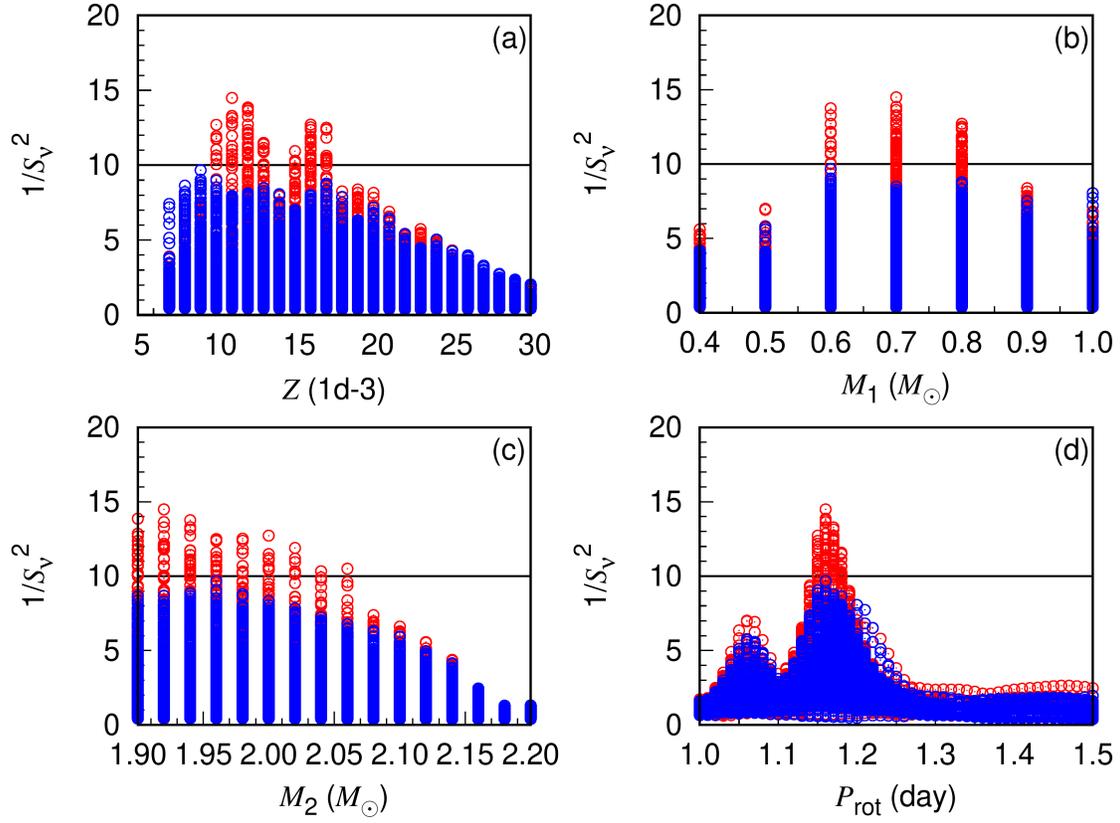}
  \caption{\label{Figure 5}  Plots of fitting results $S_{\rm m}^2$ as a function of various adjustable parameter of mass-accreting models: the metallicity $Z$,  the initial mass $M_1$,  the final mass $M_2$, and the rotation period $P_{\rm rot}$,  respectively.  The circles in red and blue correspond to accreting models of Case A and Case B evolutionary scenario,  respecitively.  The horizontal lines shows the position of $S_{\rm m}^2$ = 0.10.}
\end{figure*}
\begin{figure*}
\centering
\includegraphics[width=0.9\textwidth, angle = 0]{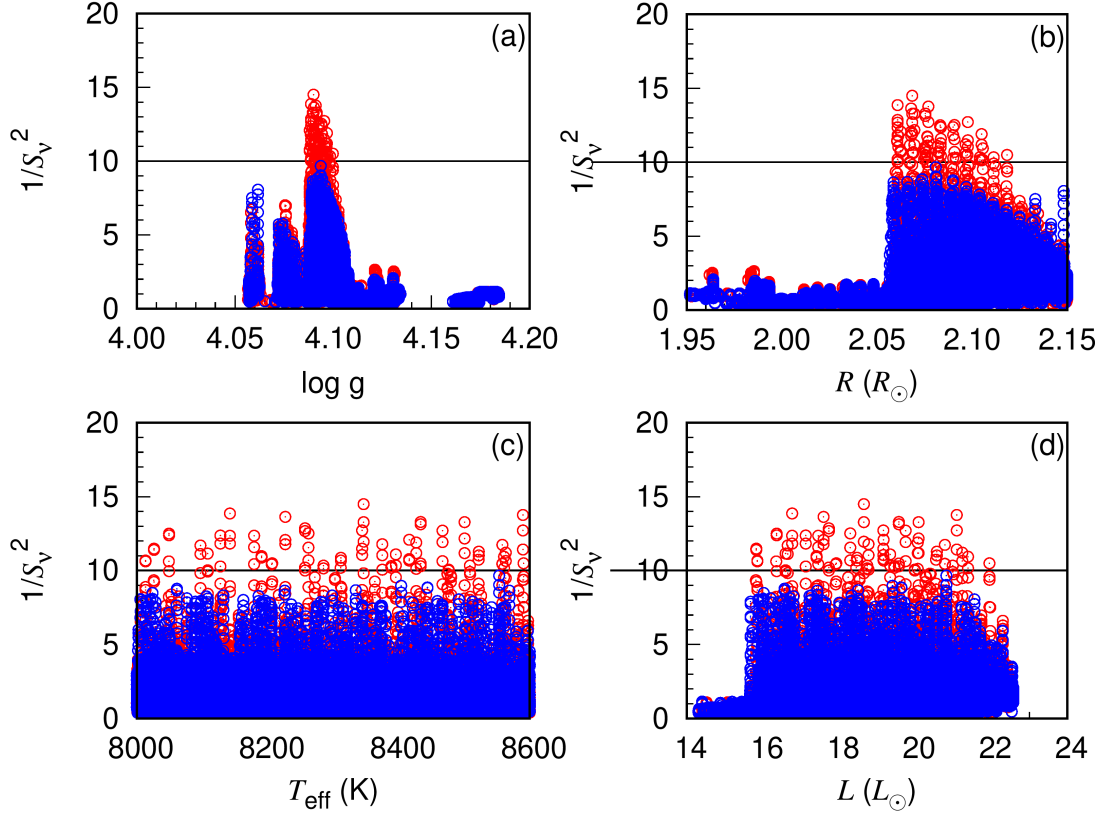}
  \caption{\label{Figure 6}  Plots of fitting results $S_{\rm m}^2$ versus stellar parameters of mass-accreting models: $\log g$,  $R$,  $T_{\rm eff}$,  and $L$,  respectively.  The circles in red and blue correspond to accreting models of Case A and Case B evolutionary scenario,  respecitively.  The horizontal lines show the position of $S_{\rm m}^2$ = 0.10.}
\end{figure*}
\begin{figure*}
\centering
\includegraphics[width=0.9\textwidth, angle = 0]{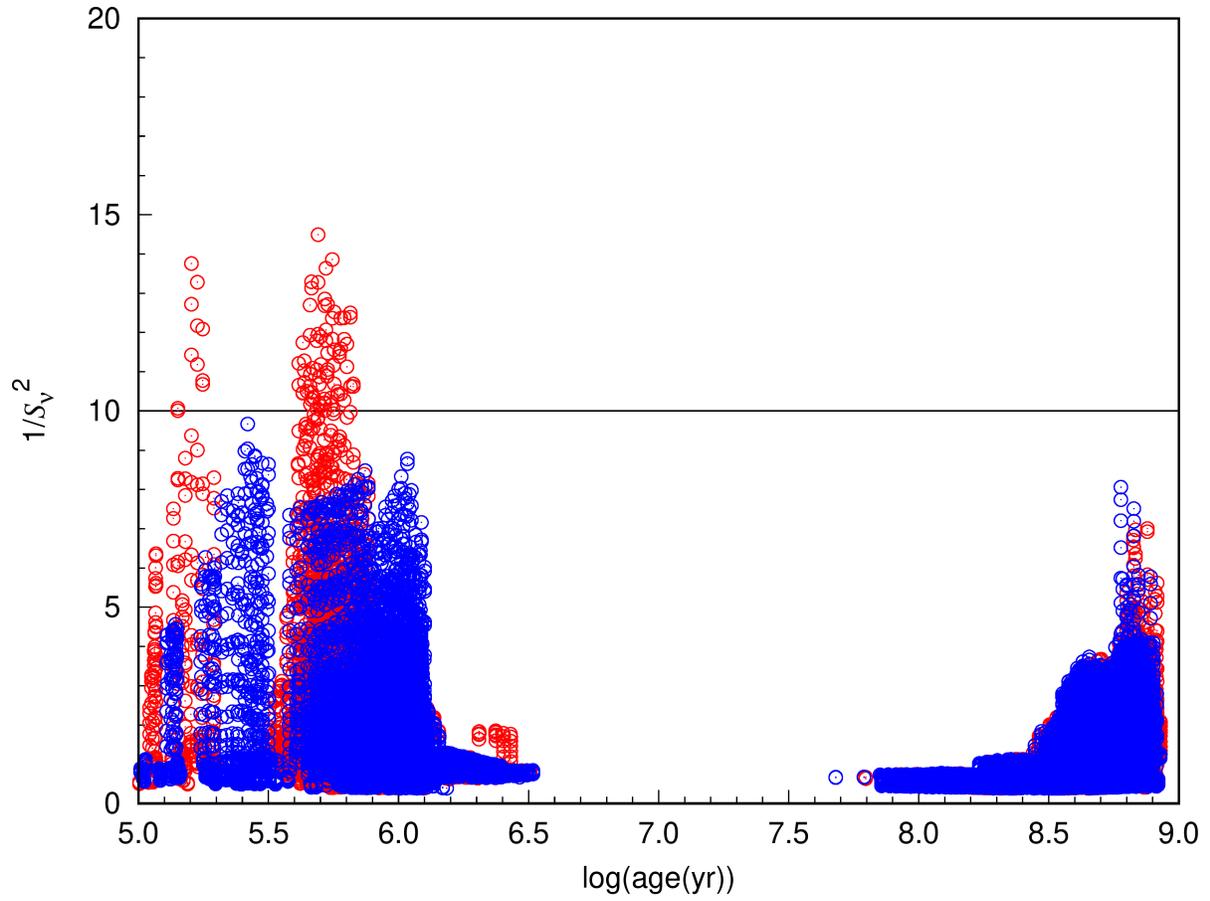}
  \caption{\label{Figure 7} Plots of fitting results $S_{\rm m}^2$ versus tge,  where tge denotes the evolutionary time of mass-accreting models since the rapid mass accretion ends. The circles in red and blue correspond to accreting models of Case A and Case B evolutionary scenario,  respecitively.  The horizontal lines show the position of $S_{\rm m}^2$ = 0.10.}
\end{figure*}
\begin{figure*}
\centering
\includegraphics[width=0.9\textwidth, angle = 0]{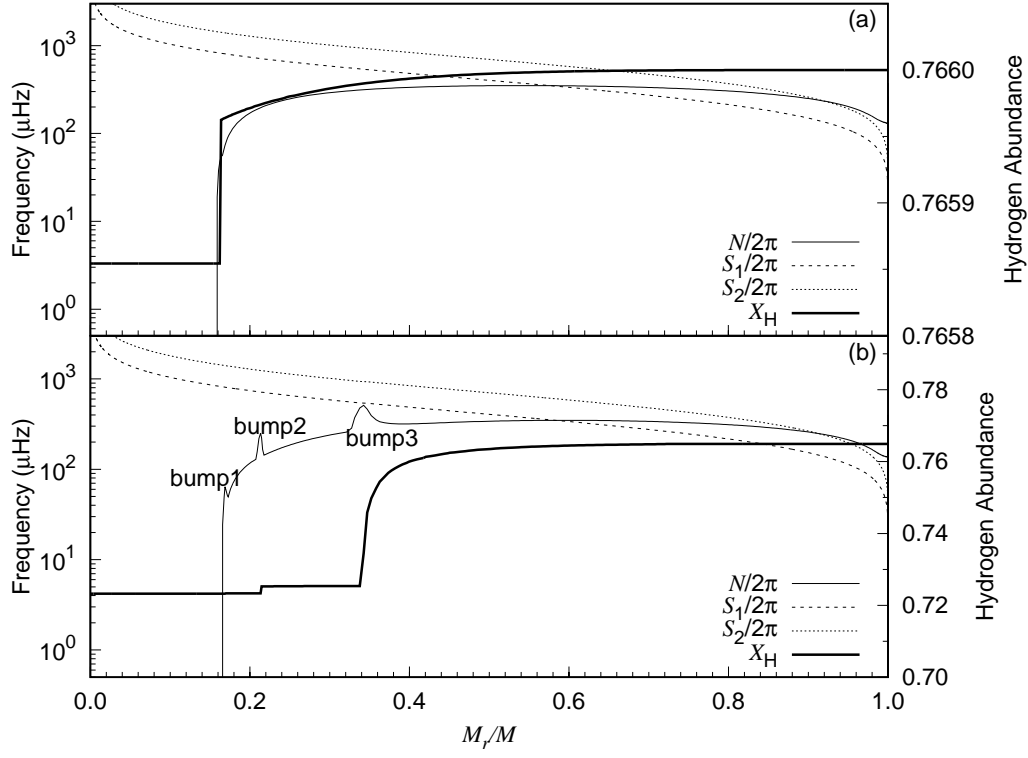}
  \caption{\label{Figure 8} Visualization of Brunt$-$V$\ddot{\rm a}$is$\ddot{\rm a}$l$\ddot{\rm a}$ frequency $N$, characteristic acoustic frequencies $S_{\ell}$ ($\ell$ = 1 and 2), and hydrogen abundance $X_{\rm H}$ inside the star.  The upper panel presents their profiles in the single-star evolutionary model (Molde A44) and the lower panel shows the profiles in mass-accreting model (Molde B21). }
\end{figure*}
\setlength{\tabcolsep}{1.2mm}{ 
\setlength{\LTcapwidth}{5in}   
\begin{table*}
\footnotesize
\centering
\caption{\label{t1}The detected frequencies of OO Dra extracted from the TESS photometric data. The columns called ID present the serial number of the detected frequencies. The columns named Freq.  show detected frequencies in units of cycle per day and $\mu$Hz,  respectively.  The columns named Ampl.  present their corresponding amplitude.  The columns named S/N present their signal-to-noise.  The remake columns show possible identifications for the detected frequencies.}
\begin{tabular}{cccccccccccc}
\hline\hline
 ID  &Freq.  &Freq.  &Ampl.  &S/N    &Remark  &ID   &Freq.   &Freq.  &Ampl.  &S/N & Remark \\
&(day$^{-1}$)   &($\mu$Hz)  &($\pm$ 24.0$\mu$mag)   & &  & &(day$^{-1}$)       &($\mu$Hz)  &($\pm$ 24.0$\mu$mag) & \\
\hline
$f_{1}$ &41.8669(4)   &484.571(5)   &2512.5  &59.2  &                         &$f_{24}$  & 0.8206(53)  &9.497(61)    & 234.2  & 5.1 &unresolved peak\\
$f_{2}$  &33.4728(5)   &387.417(6)   &2208.9  &28.0  &                         &$f_{25}$  & 1.6316(52)  &18.884(60)   & 220.8  & 6.6 &$2f_{\rm orb}$\\
$f_{3}$  &34.7407(6)   &402.092(7)   &1936.3  &37.4  &                         &$f_{26}$  &35.8564(55)  &415.004(63)  & 221.1  & 6.7 &$f_7+4f_{\rm orb}$\\
$f_{4}$  &32.4538(7)   &375.622(8)   &1633.9  &19.7  &                         &$f_{27}$  &41.2325(54)  &477.228(63)  & 215.2  & 5.9 &$f_{22}+2f_{\rm orb}$\\
$f_{5}$  &31.6572(11)  &366.402(13)  &1022.8  &25.9  &$f_4-f_{\rm orb}$        &$f_{28}$  &36.3512(58)  &420.732(67)  & 213.8  & 5.9 &$f_3+2f_{\rm orb}$\\
$f_{6}$  &33.2641(17)  &385.001(19)  & 694.0  &13.4  &$f_4+f_{\rm orb}$       &$f_{29}$  &32.5265(57)  &376.465(66)  & 203.8  & 7.3 &$f_1+f_5-f_{12}$\\
$f_{7}$  &32.6314(20)  &377.678(23)  & 583.7  &14.8  &                         &$f_{30}$  &40.2481(62)  &465.835(71)  & 191.6  & 5.4 &$f_1-2f_{\rm orb}$\\
$f_{8}$  &35.0878(25)  &406.109(28)  & 458.4  & 9.9  &$f_2+2f_{\rm orb}$       &$f_{31}$  &30.2401(63)  &350.002(73)  & 190.1  & 7.0 &$f_2-4f_{\rm orb}$\\
$f_{9}$  &33.1256(26)  &383.398(30)  & 448.6  &13.4  &$f_3-2f_{\rm orb}$       &$f_{32}$  &37.1563(66)  &430.049(77)  & 184.4  & 4.8 &$f_3+3f_{\rm orb}$\\
$f_{10}$  &34.8789(29)  &403.690(34)  & 411.2  &10.3  &$f_4+3f_{\rm orb}$       &$f_{33}$  &29.8866(73)  &345.910(84)  & 179.5  & 7.5 &$f_3-6f_{\rm orb}$\\
$f_{11}$ &38.2527(31)  &442.740(36)  & 383.9  & 9.1  &                         &$f_{34}$  & 0.7706(69)  &8.919(80)    & 175.8  & 4.4 &unresolved peak\\
$f_{12}$  &41.0008(35)  &474.546(41)  & 381.3  &10.1  &                         &$f_{35}$  & 0.8445(65)  &9.774(75)    & 183.3  & 4.9 &unresolved peak\\
$f_{13}$  &39.3842(31)  &455.836(35)  & 368.7  & 8.9  &$f_{12}-2f_{\rm orb}$    &$f_{36}$  &41.0152(81)  &474.713(94)  & 171.9  & 5.5 &unresolved peak\\
$f_{14}$  & 0.7938(33)  &9.188(38)    & 364.7  & 7.8  &$f_{\rm orb}$            &$f_{37}$  &40.6132(81)  &470.060(94)  & 150.1  & 5.0 &$f_{29}+10f_{\rm orb}$\\
$f_{15}$  &29.8652(37)  &345.662(43)  & 359.9  &10.8  &$f_2+f_{11}-f_1$          &$f_{38}$  &32.3152(80)  &374.019(93)  & 148.3  & 6.2 &$f_3-3f_{\rm orb}$\\
$f_{16}$  &28.5868(33)  &330.865(38)  & 348.3  &14.4  &$f_7-5f_{\rm orb}$       &$f_{39}$  &38.6316(82)  &447.125(94)  & 149.1  & 4.2 &$f_1-4f_{\rm orb}$\\
$f_{17}$  &34.9245(37)  &404.219(43)  & 309.5  & 8.6  &$f_3+f_7-f_4$            &$f_{40}$  &34.6681(93)  &401.251(108) & 137.6  &4.4  &$f_{23}-6f_{\rm orb}$\\
$f_{18}$  &36.5401(41)  &422.918(47)  & 288.8  & 6.0  &$f_{17}+2f_{\rm orb}$    &$f_{41}$  &36.2091(105) &419.087(121) & 129.8  & 4.3 &$f_1-7f_{\rm orb}$\\
$f_{19}$  &30.0423(43)  &347.712(49)  & 274.4  & 9.2  &$f_4-3f_{\rm orb}$       &$f_{42}$  &39.5323(123) &457.550(142) & 124.8  & 4.4 &unresolved peak\\
$f_{20}$  &37.1334(46)  &429.784(54)  & 271.4  & 5.7  &$f_{15}+9f_{\rm orb}$    &$f_{43}$  &34.5243(108) &399.587(125) & 123.9  & 4.1 &$f_{12}-8f_{\rm orb}$\\
$f_{21}$  &37.5477(43)  &434.580(50)  & 275.2  & 6.6  &$2f_8-f_7$               &$f_{44}$  &41.4923(104) &480.236(121) & 119.3  & 4.0 &$f_{11}+4f_{\rm orb}$\\
$f_{22}$ &39.6148(42)  &458.504(49)  & 255.0  & 6.7  &$f_1+f_7-f_{10}$         &$f_{45}$  &35.4024(112) &409.750(129) & 117.6  & 4.2 &$f_1-8f_{\rm orb}$\\
$f_{23}$  &39.5176(55)  &457.379(63)  & 249.7  & 7.3  &$f_3+f_{11}-f_2$\\
\hline
\end{tabular}
\end{table*}

\setlength{\tabcolsep}{3.6mm}{ 
\setlength{\LTcapwidth}{8in}   
\begin{table*}
\centering
\scriptsize
\caption{\label{t2}Preferred models with $S_{\rm m}^{2}$ $\le$ 0.10. $P_{\rm rot}$ represents the rotation period,  $\tau_0$ represnets the acoustic radius,  and $X_{\rm c}$ represents the mass fraction of hydrogen in the center of the star. }
\begin{tabular}{ccccccccccccccccc}
\hline\hline 
Model &$P_{\rm rot}$ &$Z$  &$M$  &$Y$ &$T_{\rm eff}$ &log$g$   &$R$ &$L$ &$\tau_0$ &$X_{\rm c}$ &Age &$S_{\rm m}^{2}$  \\
&(day)  &     &($M_{\odot}$)  & &(K)   &(cgs) &$(R_{\odot})$  &($L_{\odot}$) &(hr)  & &(Myr)  &  \\
\hline
A01  &1.15  &0.013  &1.94  &0.220  &8172  &4.091  &2.076  &17.27  &2.308  &0.767  &7.93  &0.098\\
A02  &1.15  &0.013  &1.96  &0.220  &8241  &4.093  &2.083  &17.98  &2.311  &0.767  &7.75  &0.094\\
A03  &1.15  &0.013  &2.00  &0.220  &8376  &4.096  &2.097  &19.44  &2.317  &0.767  &7.41  &0.099\\
A04  &1.15  &0.014  &1.94  &0.220  &8065  &4.091  &2.076  &16.38  &2.305  &0.766  &8.22  &0.090\\
A05  &1.15  &0.014  &1.96  &0.220  &8131  &4.093  &2.083  &17.04  &2.307  &0.766  &8.04  &0.088\\
A06  &1.15  &0.015  &1.96  &0.220  &8020  &4.093  &2.082  &16.12  &2.302  &0.765  &8.34  &0.097\\
A07  &1.16  &0.011  &2.00  &0.220  &8593  &4.096  &2.097  &21.54  &2.324  &0.769  &6.88  &0.098\\
A08  &1.16  &0.012  &1.94  &0.220  &8277  &4.091  &2.076  &18.18  &2.313  &0.768  &7.64  &0.094\\
A09  &1.16  &0.012  &1.96  &0.220  &8347  &4.093  &2.083  &18.93  &2.316  &0.768  &7.47  &0.090\\
A10  &1.16  &0.012  &1.98  &0.220  &8417  &4.094  &2.090  &19.69  &2.317  &0.768  &7.30  &0.091\\
A11  &1.16  &0.012  &2.00  &0.220  &8486  &4.096  &2.097  &20.48  &2.320  &0.768  &7.14  &0.082\\
A12  &1.16  &0.012  &2.02  &0.220  &8554  &4.097  &2.104  &21.30  &2.322  &0.768  &6.99  &0.090\\
A13  &1.16  &0.013  &1.90  &0.220  &8034  &4.088  &2.062  &15.91  &2.303  &0.767  &8.30  &0.095\\
A14  &1.16  &0.013  &1.92  &0.220  &8103  &4.090  &2.069  &16.58  &2.305  &0.767  &8.11  &0.081\\
A15  &1.16  &0.013  &1.94  &0.220  &8172  &4.091  &2.076  &17.27  &2.308  &0.767  &7.93  &0.077\\
A16  &1.16  &0.013  &1.96  &0.220  &8241  &4.093  &2.083  &17.98  &2.311  &0.767  &7.75  &0.076\\
A17  &1.16  &0.013  &1.98  &0.220  &8309  &4.094  &2.090  &18.70  &2.313  &0.767  &7.58  &0.079\\
A18  &1.16  &0.013  &2.00  &0.220  &8376  &4.096  &2.097  &19.44  &2.317  &0.767  &7.41  &0.084\\
A19  &1.16  &0.013  &2.02  &0.220  &8443  &4.097  &2.104  &20.20  &2.318  &0.767  &7.25  &0.090\\
A20  &1.16  &0.013  &2.04  &0.220  &8509  &4.099  &2.110  &20.98  &2.321  &0.767  &7.09  &0.095\\
A21  &1.16  &0.014  &1.94  &0.220  &8065  &4.092  &2.075  &16.37  &2.304  &0.766  &8.22  &0.074\\
A22  &1.16  &0.014  &1.96  &0.220  &8131  &4.093  &2.083  &17.04  &2.307  &0.766  &8.04  &0.076\\
A23  &1.16  &0.014  &1.98  &0.220  &8198  &4.095  &2.089  &17.71  &2.309  &0.766  &7.86  &0.083\\
A24  &1.16  &0.014  &2.00  &0.220  &8263  &4.096  &2.097  &18.41  &2.313  &0.766  &7.69  &0.096\\
A25  &1.16  &0.015  &1.96  &0.220  &8020  &4.093  &2.082  &16.11  &2.301  &0.765  &8.34  &0.088\\
A26  &1.17  &0.011  &1.94  &0.220  &8379  &4.091  &2.076  &19.08  &2.317  &0.769  &7.36  &0.096\\
A27  &1.17  &0.011  &1.96  &0.220  &8451  &4.093  &2.083  &19.88  &2.318  &0.769  &7.20  &0.088\\
A28  &1.17  &0.011  &1.98  &0.220  &8522  &4.094  &2.090  &20.70  &2.321  &0.769  &7.04  &0.086\\
A29  &1.17  &0.011  &2.00  &0.220  &8593  &4.096  &2.097  &21.54  &2.324  &0.769  &6.88  &0.083\\
A30  &1.17  &0.012  &1.90  &0.220  &8135  &4.088  &2.062  &16.73  &2.306  &0.768  &8.00  &0.099\\
A31  &1.17  &0.012  &1.92  &0.220  &8206  &4.090  &2.069  &17.44  &2.309  &0.768  &7.82  &0.092\\
A32  &1.17  &0.012  &1.94  &0.220  &8278  &4.091  &2.075  &18.17  &2.312  &0.768  &7.64  &0.089\\
A33  &1.17  &0.012  &1.96  &0.220  &8348  &4.093  &2.083  &18.92  &2.315  &0.768  &7.47  &0.085\\
A34  &1.17  &0.012  &1.98  &0.220  &8417  &4.094  &2.090  &19.69  &2.317  &0.768  &7.30  &0.079\\
A35  &1.17  &0.012  &2.00  &0.220  &8486  &4.096  &2.097  &20.48  &2.320  &0.768  &7.14  &0.076\\
A36  &1.17  &0.012  &2.02  &0.220  &8554  &4.097  &2.104  &21.30  &2.322  &0.768  &6.99  &0.092\\
A37  &1.17  &0.013  &1.90  &0.220  &8034  &4.089  &2.061  &15.90  &2.302  &0.767  &8.30  &0.083\\
A38  &1.17  &0.013  &1.92  &0.220  &8103  &4.090  &2.069  &16.58  &2.305  &0.767  &8.11  &0.078\\
A39  &1.17  &0.013  &1.94  &0.220  &8172  &4.091  &2.076  &17.27  &2.308  &0.767  &7.93  &0.078\\
A40  &1.17  &0.013  &1.96  &0.220  &8241  &4.093  &2.083  &17.98  &2.311  &0.767  &7.75  &0.079\\
A41  &1.17  &0.013  &1.98  &0.220  &8309  &4.094  &2.090  &18.70  &2.313  &0.767  &7.58  &0.075\\
A42  &1.17  &0.013  &2.00  &0.220  &8376  &4.096  &2.097  &19.44  &2.317  &0.767  &7.41  &0.091\\
A43  &1.17  &0.013  &2.02  &0.220  &8443  &4.097  &2.104  &20.20  &2.318  &0.767  &7.25  &0.098\\
A44  &1.17  &0.014  &1.94  &0.220  &8065  &4.092  &2.075  &16.37  &2.304  &0.766  &8.22  &0.072\\
A45  &1.17  &0.014  &1.96  &0.220  &8132  &4.093  &2.082  &17.03  &2.306  &0.766  &8.04  &0.085\\
A46  &1.17  &0.014  &1.98  &0.220  &8198  &4.095  &2.089  &17.71  &2.309  &0.766  &7.86  &0.086\\
A47  &1.17  &0.015  &1.96  &0.220  &8020  &4.093  &2.082  &16.11  &2.301  &0.765  &8.34  &0.095\\
A48  &1.18  &0.011  &1.96  &0.220  &8451  &4.093  &2.083  &19.88  &2.318  &0.769  &7.20  &0.094\\
A49  &1.18  &0.011  &1.98  &0.220  &8522  &4.094  &2.090  &20.70  &2.321  &0.769  &7.04  &0.088\\
A50  &1.18  &0.011  &2.00  &0.220  &8593  &4.096  &2.097  &21.54  &2.324  &0.769  &6.88  &0.088\\
A51  &1.18  &0.012  &1.92  &0.220  &8207  &4.090  &2.068  &17.44  &2.308  &0.768  &7.82  &0.098\\
A52  &1.18  &0.012  &1.94  &0.220  &8278  &4.091  &2.075  &18.17  &2.312  &0.768  &7.64  &0.091\\
A53  &1.18  &0.012  &1.96  &0.220  &8348  &4.093  &2.083  &18.92  &2.315  &0.768  &7.47  &0.088\\
A54  &1.18  &0.012  &1.98  &0.220  &8417  &4.094  &2.090  &19.69  &2.317  &0.768  &7.30  &0.087\\
A55  &1.18  &0.012  &2.00  &0.220  &8486  &4.096  &2.097  &20.48  &2.320  &0.768  &7.14  &0.089\\
A56  &1.18  &0.013  &1.90  &0.220  &8034  &4.089  &2.061  &15.90  &2.302  &0.767  &8.30  &0.089\\
A57  &1.18  &0.013  &1.92  &0.220  &8104  &4.090  &2.068  &16.57  &2.304  &0.767  &8.11  &0.090\\
A58  &1.18  &0.013  &1.94  &0.220  &8173  &4.092  &2.075  &17.26  &2.307  &0.767  &7.93  &0.090\\
A59  &1.18  &0.013  &1.96  &0.220  &8241  &4.093  &2.082  &17.97  &2.310  &0.767  &7.75  &0.092\\
A60  &1.18  &0.013  &1.98  &0.220  &8309  &4.094  &2.090  &18.70  &2.313  &0.767  &7.58  &0.091\\
A61  &1.18  &0.014  &1.94  &0.220  &8065  &4.092  &2.075  &16.37  &2.304  &0.766  &8.22  &0.090\\
\hline
\end{tabular}
\end{table*}
\setlength{\tabcolsep}{5mm}{ 
\setlength{\LTcapwidth}{7in}   
\begin{table*}
\centering
\caption{\label{t3}Fundamental parameters of the primary component of OO Dra yielded by asteroseismology.  $P_{\rm rot}$ is the rotation period. $\tau_0$ represents the acoustic radius.  $X_{\rm c}$ denotes the mass fraction of hydrogen in the center of the star.  Tge denotes the evolutionary time of mass-accreting models since the rapid mass accretion ends.  Stellar parameters of Lee et al.  (2018) by the binary model are listed in the last column. }
\begin{tabular}{lllllll}
\hline\hline
Parameters                   &Single-star models                                  &Mass-accreting models               &Lee et al. (2018)\\
\hline
$Z$                                &0.011$-$0.015 (0.014$_{-0.003}^{+0.001}$)            &0.010-0.017 (0.011$_{-0.001}^{+0.006}$)   &\\
$M$ ($M_{\odot}$)      &1.90$-$2.04 (1.94$_{-0.04}^{+0.10}$)                      &\nodata                                                                 &2.031$\pm$0.058\\
$Y$                                 &0.220                                                                              &0.249+1.33$Z$\\
$M_1$ ($M_{\odot}$)   &\nodata                                                                           &0.60-0.80 (0.70$\pm$ 0.10)\\
$M_2$ ($M_{\odot}$)  &\nodata                                                                          &1.90-2.02 (1.92$_{-0.02}^{+0.10}$)\\
$P_{\rm rot}$(day)       &1.15-1.18 ($1.17^{+0.01}_{-0.02}$)                          &1.15-1.18 (1.16$_{-0.01}^{+0.02}$)\\
$T_{\rm eff}$ (K)          &8020-8593 ($8065^{+528}_{-45}$)                      &8026-8590 (8346$_{-320}^{+244}$)              &8260$\pm$210\\
$\log g$ (cgs)               &4.088-4.099 ($4.092^{+0.007}_{-0.004}$)          &4.088-4.100 (4.090$_{-0.002}^{+0.010}$)     &4.110$\pm$0.017\\
$R$ ($R_{\odot}$)        &2.061-2.110 ($2.075^{+0.035}_{-0.014}$)            &2.061-2.118 (2.068$_{-0.007}^{+0.050}$)       &2.078$\pm$0.026\\
$L$ ($L_{\odot}$)         &15.90-21.54 ($16.37^{+5.17}_{-0.47}$)                &15.83-21.96 (18.65$_{-2.82}^{+3.31}$)            &18.0$\pm$1.9\\
$\tau_0$ (hr)                   &2.301-2.324 ($2.304^{+0.020}_{-0.003}$)       &2.302-2.324 (2.315$_{-0.013}^{+0.009}$)\\
$X_{\rm c}$                   &0.765-0.769 (0.766$^{+0.003}_{-0.001}$)         &0.696-0.726 (0.723$^{+0.003}_{-0.027}$)\\
Age (Myr)                       &6.89-8.34 ($8.22^{+0.12}_{-1.33}$)     \\
Tge (Myr)                       &                                                                                    &0.14-0.67 (0.49$^{+0.18}_{-0.35}$)\\
\hline
\end{tabular}
\end{table*}
\begin{table*}
\centering
\caption{\label{t4}Theoretical frequencies of the best-fitting single-star evolutionary model (Model A44).  $\nu_{\rm mod}$ is the model frequency.  $\ell$ and $n$ are its spherical harmonic degree and radial order, respectively. $\beta_{\ell,n}$ is the rotational parameters defined as equation (2). }
\label{observed frequencies}
\begin{tabular}{cccccc}
\hline\hline
$\nu_{\rm mod}(\ell,n)$ &$\beta_{\ell, n}$ &$\nu_{\rm mod}(\ell,n)$ &$\beta_{\ell, n}$ &$\nu_{\rm mod}(\ell,n)$ &$\beta_{\ell, n}$\\
     ($\mu$Hz)    &     &($\mu$Hz)              &              &($\mu$Hz)\\
\hline
166.800(0,0)  &  &124.265(1,0)  &0.487  &164.889(2,0)  &0.918\\
215.551(0,1)  &  &171.134(1,1)  &0.985  &191.521(2,0)  &0.861\\
265.700(0,2)  &  &222.486(1,2)  &0.997  &226.527(2,1)  &0.996\\
315.399(0,3)  &  &276.840(1,3)  &0.998  &261.773(2,2)  &0.878\\
365.904(0,4)  &  &332.122(1,4)  &0.996  &305.360(2,3)  &0.879\\
417.999(0,5)  &  &387.732(1,5)  &0.993  &357.115(2,4)  &0.925\\
470.551(0,6)  &  &442.375(1,6)  &0.990  &411.472(2,5)  &0.951\\
523.603(0,7)  &  &496.263(1,7)  &0.989  &465.420(2,6)  &0.964\\
577.884(0,8)  &  &550.456(1,8)  &0.988  &519.169(2,7)  &0.972\\
633.391(0,9)  &  &605.631(1,9)  &0.987  &573.737(2,8)  &0.977\\
689.930(0,10) &  &661.736(1,10) &0.986  &629.321(2,9)  &0.980\\
\hline
\end{tabular}
\end{table*}
\begin{table*}
\centering
\caption{\label{t5}Comparisons between theoretical frequencies of the best-fitting single-star evolutionary model (Model A44) and observations. $\nu_{\rm obs}$ denotes the observed frequency.  $\nu_{\rm mod}$ represents its model counterpart.  $|\nu_{\rm obs}-\nu_{\rm mod}|$ shows the difference between the observed frequency and the model counterpart.}
\begin{tabular}{ccclc}
\hline\hline
ID  &$\nu_{\rm obs}$   &$\nu_{\rm mod}$ &($\ell$, $n$, $m$) & $|\nu_{\rm obs}-\nu_{\rm mod}|$\\
     &($\mu$Hz)         &($\mu$Hz)              &              &($\mu$Hz)\\
\hline
$f_{1}$      &484.571           &484.493         &(2,  6,  +2)        &0.078\\
$f_{2}$     &387.417            &387.732          &(1,  5,  0)          &0.315\\
$f_{3}$     &402.092          &402.064         &(2,  5,  -1)         &0.028\\
$f_{4}$     &375.622          &375.416          &(2,  4,  +2)        &0.206\\
$f_{7}$     &377.678           &377.909         &(1,  5,  -1)          &0.231\\
$f_{11}$    &442.740          &442.375        &(1,  6,  0)           &0.365\\
$f_{12}$    &474.546          &474.957        &(2,  6,  +1)         &0.411\\
\hline
\end{tabular}
\end{table*}
\setlength{\tabcolsep}{3.2mm}{ 
\setlength{\LTcapwidth}{6in}   
\begin{table*}
\centering
\caption{\label{t6}Preferred mass-accreting models with $S_{\rm m}^{2}$ $\le$ 0.10. $P_{\rm rot}$ represents the rotation period.  $\tau_0$ represents the acoustic radius.  $X_{\rm c}$ represents the mass fraction of hydrogen in the center of the star.  Tge denotes the evolutionary time of mass-accreting models since the rapid mass accretion ends.}
\begin{tabular}{ccccccccccccc}
\hline\hline
Model &$P_{\rm rot}$ &$Z$  &$M_1$   &$M_2$   &$T_{\rm eff}$ &log$g$    &$R$  &$L$ &$\tau_0$ &$X_{\rm c}$ &Tge &$S_{\rm m}^{2}$ \\
&(day)  &     &($M_{\odot}$) &($M_{\odot}$) &(K) &  &$(R_{\odot})$  & ($L_{\odot}$) &(hr)   & &(Myr)  & \\
\hline
B01  &1.15  &0.010  &0.70  &1.94  &8558  &4.091  &2.076  &20.78  &2.322  &0.726  &0.43  &0.096\\
B02  &1.15  &0.011  &0.70  &1.90  &8257  &4.089  &2.061  &17.74  &2.311  &0.723  &0.52  &0.098\\
B03  &1.15  &0.011  &0.70  &1.92  &8346  &4.090  &2.068  &18.65  &2.315  &0.723  &0.49  &0.084\\
B04  &1.15  &0.011  &0.70  &1.94  &8433  &4.092  &2.075  &19.57  &2.318  &0.723  &0.46  &0.095\\
B05  &1.15  &0.011  &0.70  &1.96  &8522  &4.093  &2.083  &20.57  &2.322  &0.723  &0.43  &0.091\\
B06  &1.15  &0.012  &0.60  &1.90  &8416  &4.088  &2.062  &19.17  &2.316  &0.722  &0.18  &0.094\\
B07  &1.15  &0.012  &0.60  &1.92  &8501  &4.090  &2.070  &20.09  &2.319  &0.722  &0.17  &0.082\\
B08  &1.15  &0.012  &0.60  &1.94  &8590  &4.091  &2.077  &21.10  &2.322  &0.722  &0.16  &0.079\\
B09  &1.15  &0.012  &0.70  &1.90  &8142  &4.089  &2.061  &16.77  &2.306  &0.721  &0.56  &0.081\\
B10  &1.15  &0.012  &0.70  &1.92  &8227  &4.090  &2.068  &17.60  &2.309  &0.721  &0.53  &0.085\\
B11  &1.15  &0.012  &0.70  &1.94  &8312  &4.091  &2.076  &18.47  &2.313  &0.721  &0.50  &0.092\\
B12  &1.15  &0.012  &0.70  &2.00  &8563  &4.096  &2.097  &21.24  &2.322  &0.721  &0.41  &0.094\\
B13  &1.15  &0.013  &0.60  &1.94  &8476  &4.091  &2.076  &19.99  &2.318  &0.720  &0.14  &0.099\\
B14  &1.15  &0.013  &0.70  &1.90  &8026  &4.089  &2.061  &15.83  &2.302  &0.719  &0.59  &0.087\\
B15  &1.15  &0.013  &0.70  &1.92  &8108  &4.090  &2.067  &16.60  &2.305  &0.719  &0.56  &0.100\\
B16  &1.15  &0.013  &0.70  &1.94  &8191  &4.092  &2.075  &17.41  &2.309  &0.719  &0.53  &0.091\\
B17  &1.16  &0.010  &0.70  &1.90  &8375  &4.089  &2.061  &18.78  &2.314  &0.726  &0.49  &0.091\\
B18  &1.16  &0.010  &0.70  &1.92  &8467  &4.090  &2.068  &19.75  &2.318  &0.726  &0.46  &0.084\\
B19  &1.16  &0.010  &0.70  &1.94  &8558  &4.091  &2.076  &20.78  &2.322  &0.726  &0.43  &0.085\\
B20  &1.16  &0.011  &0.70  &1.90  &8257  &4.089  &2.061  &17.74  &2.311  &0.723  &0.52  &0.078\\
B21  &1.16  &0.011  &0.70  &1.92  &8346  &4.090  &2.068  &18.65  &2.315  &0.723  &0.49  &0.069\\
B22  &1.16  &0.011  &0.70  &1.94  &8433  &4.092  &2.075  &19.57  &2.318  &0.723  &0.46  &0.075\\
B23  &1.16  &0.011  &0.70  &1.96  &8522  &4.093  &2.083  &20.57  &2.322  &0.723  &0.43  &0.089\\
B24  &1.16  &0.012  &0.60  &1.90  &8416  &4.088  &2.062  &19.17  &2.316  &0.722  &0.18  &0.083\\
B25  &1.16  &0.012  &0.60  &1.92  &8501  &4.090  &2.070  &20.09  &2.319  &0.722  &0.17  &0.075\\
B26  &1.16  &0.012  &0.60  &1.94  &8590  &4.091  &2.077  &21.10  &2.322  &0.722  &0.16  &0.073\\
B27  &1.16  &0.012  &0.70  &1.90  &8142  &4.089  &2.061  &16.77  &2.306  &0.721  &0.56  &0.072\\
B28  &1.16  &0.012  &0.70  &1.92  &8227  &4.090  &2.068  &17.60  &2.309  &0.721  &0.53  &0.073\\
B29  &1.16  &0.012  &0.70  &1.94  &8312  &4.091  &2.076  &18.47  &2.313  &0.721  &0.50  &0.094\\
B30  &1.16  &0.012  &0.70  &1.96  &8396  &4.093  &2.082  &19.35  &2.315  &0.721  &0.47  &0.090\\
B31  &1.16  &0.012  &0.70  &2.00  &8563  &4.096  &2.097  &21.24  &2.322  &0.721  &0.41  &0.089\\
B32  &1.16  &0.013  &0.60  &1.94  &8476  &4.091  &2.076  &19.99  &2.318  &0.720  &0.14  &0.100\\
B33  &1.16  &0.013  &0.70  &1.90  &8026  &4.089  &2.061  &15.83  &2.302  &0.719  &0.59  &0.088\\
B34  &1.16  &0.013  &0.70  &1.92  &8108  &4.090  &2.067  &16.60  &2.305  &0.719  &0.56  &0.090\\
B35  &1.16  &0.013  &0.70  &1.94  &8191  &4.092  &2.075  &17.41  &2.309  &0.719  &0.53  &0.092\\
B36  &1.16  &0.016  &0.80  &1.96  &8179  &4.093  &2.084  &17.45  &2.309  &0.698  &0.60  &0.096\\
B37  &1.16  &0.016  &0.80  &1.98  &8262  &4.094  &2.091  &18.30  &2.312  &0.698  &0.57  &0.094\\
B38  &1.16  &0.016  &0.80  &2.00  &8344  &4.095  &2.098  &19.16  &2.315  &0.698  &0.53  &0.090\\
B39  &1.16  &0.016  &0.80  &2.02  &8427  &4.097  &2.105  &20.07  &2.318  &0.698  &0.50  &0.089\\
B40  &1.16  &0.017  &0.80  &1.96  &8049  &4.093  &2.083  &16.37  &2.303  &0.696  &0.65  &0.080\\
B41  &1.16  &0.017  &0.80  &1.98  &8128  &4.094  &2.090  &17.13  &2.306  &0.696  &0.62  &0.085\\
B42  &1.16  &0.017  &0.80  &2.00  &8207  &4.096  &2.097  &17.93  &2.309  &0.696  &0.58  &0.095\\
B43  &1.16  &0.017  &0.80  &2.02  &8285  &4.097  &2.104  &18.74  &2.312  &0.696  &0.55  &0.097\\
B44  &1.17  &0.010  &0.70  &1.90  &8375  &4.089  &2.061  &18.78  &2.314  &0.726  &0.49  &0.085\\
B45  &1.17  &0.010  &0.70  &1.92  &8467  &4.090  &2.068  &19.75  &2.318  &0.726  &0.46  &0.079\\
B46  &1.17  &0.010  &0.70  &1.94  &8558  &4.091  &2.076  &20.78  &2.322  &0.726  &0.43  &0.096\\
B47  &1.17  &0.011  &0.70  &1.90  &8257  &4.089  &2.061  &17.74  &2.311  &0.723  &0.52  &0.079\\
B48  &1.17  &0.011  &0.70  &1.92  &8346  &4.090  &2.068  &18.65  &2.315  &0.723  &0.49  &0.075\\
B49  &1.17  &0.011  &0.70  &1.94  &8433  &4.092  &2.075  &19.57  &2.318  &0.723  &0.46  &0.076\\
B50  &1.17  &0.011  &0.70  &1.96  &8522  &4.093  &2.082  &20.55  &2.320  &0.723  &0.44  &0.093\\
\hline
\end{tabular}
\end{table*}
\begin{table*}
\centering
\caption{\label{t7}Comparisons between model frequencies and observations for the best-fitting model (Model B21) in mass-accreting evolutionary grid.  $\nu_{\rm obs}$ denotes the observed frequency.  $\nu_{\rm mod}$ denotes the model frequency.  $|\nu_{\rm obs}-\nu_{\rm mod}|$ is the difference between the observed frequency and its model counterpart.}
\begin{tabular}{ccclc}
\hline\hline
ID  &$\nu_{\rm obs}$   &$\nu_{\rm mod}$ &($\ell$, $n$, $m$) & $|\nu_{\rm obs}-\nu_{\rm mod}|$\\
     &($\mu$Hz)         &($\mu$Hz)              &              &($\mu$Hz)\\
\hline
$f_{1}$      &484.571           &484.464         &(2,  6,  +2)        &0.107\\
$f_{2}$     &387.417            &387.756          &(1,  5,  0)          &0.339\\
$f_{3}$     &402.092          &401.898         &(2,  5,  -1)         &0.194\\
$f_{4}$     &375.622          &375.656          &(2,  4,  +2)        &0.034\\
$f_{7}$     &377.678           &377.848          &(1,  5,  -1)          &0.170\\
$f_{11}$    &442.740          &442.294        &(1,  6,  0)           &0.446\\
$f_{12}$    &474.546          &474.846        &(2,  6,  +1)         &0.300\\
\hline
\end{tabular}
\end{table*}
\setlength{\tabcolsep}{3.2mm}{ 
\setlength{\LTcapwidth}{6in}   
\begin{table*}
\centering
\caption{\label{t6-continue} Table 6 - continued}
\begin{tabular}{ccccccccccccccccc}
\hline\hline
Model &$P_{\rm rot}$ &$Z$  &$M_1$   &$M_2$   &$T_{\rm eff}$ &log$g$    &$R$  &$L$ &$\tau_0$ &$X_{\rm c}$ &Tge &$S_{\rm m}^{2}$ \\
&(day)  &     &($M_{\odot}$) &($M_{\odot}$) &(K) &  &$(R_{\odot})$  & ($L_{\odot}$) &(hr)   & &(Myr)  & \\
\hline
B51  &1.17  &0.012  &0.60  &1.90  &8416  &4.088  &2.062  &19.17  &2.316  &0.722  &0.18  &0.093\\
B52  &1.17  &0.012  &0.60  &1.92  &8501  &4.090  &2.070  &20.09  &2.319  &0.722  &0.17  &0.089\\
B53  &1.17  &0.012  &0.60  &1.94  &8590  &4.091  &2.077  &21.10  &2.322  &0.722  &0.16  &0.088\\
B54  &1.17  &0.012  &0.70  &1.90  &8142  &4.089  &2.061  &16.77  &2.306  &0.721  &0.56  &0.085\\
B55  &1.17  &0.012  &0.70  &1.92  &8227  &4.090  &2.068  &17.60  &2.309  &0.721  &0.53  &0.083\\
B56  &1.17  &0.012  &0.70  &1.94  &8311  &4.092  &2.075  &18.45  &2.312  &0.721  &0.50  &0.098\\
B57  &1.17  &0.012  &0.70  &1.96  &8396  &4.093  &2.082  &19.35  &2.315  &0.721  &0.47  &0.097\\
B58  &1.17  &0.015  &0.80  &1.98  &8389  &4.094  &2.091  &19.46  &2.316  &0.700  &0.52  &0.097\\
B59  &1.17  &0.015  &0.80  &2.02  &8561  &4.097  &2.106  &21.39  &2.323  &0.700  &0.46  &0.091\\
B60  &1.17  &0.016  &0.80  &1.92  &8013  &4.090  &2.069  &15.86  &2.302  &0.698  &0.67  &0.094\\
B61  &1.17  &0.016  &0.80  &1.94  &8097  &4.091  &2.076  &16.65  &2.305  &0.698  &0.63  &0.085\\
B62  &1.17  &0.016  &0.80  &1.96  &8179  &4.093  &2.084  &17.45  &2.309  &0.698  &0.60  &0.081\\
B63  &1.17  &0.016  &0.80  &1.98  &8262  &4.094  &2.091  &18.30  &2.312  &0.698  &0.57  &0.080\\
B64  &1.17  &0.016  &0.80  &2.00  &8344  &4.095  &2.098  &19.16  &2.315  &0.698  &0.53  &0.079\\
B65  &1.17  &0.016  &0.80  &2.02  &8427  &4.097  &2.105  &20.07  &2.318  &0.698  &0.50  &0.084\\
B66  &1.17  &0.016  &0.80  &2.04  &8509  &4.098  &2.111  &20.99  &2.320  &0.698  &0.47  &0.097\\
B67  &1.17  &0.016  &0.80  &2.06  &8590  &4.100  &2.118  &21.96  &2.324  &0.698  &0.45  &0.095\\
B68  &1.17  &0.017  &0.80  &1.96  &8049  &4.093  &2.083  &16.37  &2.303  &0.696  &0.65  &0.081\\
B69  &1.17  &0.017  &0.80  &1.98  &8128  &4.094  &2.090  &17.13  &2.306  &0.696  &0.62  &0.081\\
B70  &1.17  &0.017  &0.80  &2.00  &8206  &4.096  &2.097  &17.91  &2.308  &0.696  &0.58  &0.096\\
B71  &1.17  &0.017  &0.80  &2.02  &8285  &4.097  &2.104  &18.74  &2.312  &0.696  &0.55  &0.099\\
B72  &1.18  &0.010  &0.70  &1.90  &8375  &4.089  &2.061  &18.78  &2.314  &0.726  &0.49  &0.099\\
B73  &1.18  &0.010  &0.70  &1.92  &8467  &4.090  &2.068  &19.75  &2.318  &0.726  &0.46  &0.094\\
B74  &1.18  &0.011  &0.70  &1.90  &8257  &4.089  &2.061  &17.74  &2.311  &0.723  &0.52  &0.100\\
B75  &1.18  &0.011  &0.70  &1.94  &8433  &4.092  &2.075  &19.57  &2.318  &0.723  &0.46  &0.097\\
B76  &1.18  &0.015  &0.80  &1.98  &8389  &4.094  &2.091  &19.46  &2.316  &0.700  &0.52  &0.098\\
B77  &1.18  &0.015  &0.80  &2.02  &8561  &4.097  &2.106  &21.39  &2.323  &0.700  &0.46  &0.098\\
B78  &1.18  &0.016  &0.80  &1.92  &8013  &4.090  &2.069  &15.86  &2.302  &0.698  &0.67  &0.094\\
B79  &1.18  &0.016  &0.80  &1.94  &8097  &4.091  &2.076  &16.65  &2.305  &0.698  &0.63  &0.090\\
B80  &1.18  &0.016  &0.80  &1.96  &8179  &4.093  &2.084  &17.45  &2.309  &0.698  &0.60  &0.086\\
B81  &1.18  &0.016  &0.80  &1.98  &8262  &4.094  &2.091  &18.30  &2.312  &0.698  &0.57  &0.087\\
B82  &1.18  &0.016  &0.80  &2.00  &8344  &4.095  &2.098  &19.16  &2.315  &0.698  &0.53  &0.087\\
B83  &1.18  &0.016  &0.80  &2.02  &8427  &4.097  &2.105  &20.07  &2.318  &0.698  &0.50  &0.099\\
B84  &1.18  &0.017  &0.80  &1.96  &8049  &4.093  &2.083  &16.35  &2.302  &0.696  &0.65  &0.094\\
B85  &1.18  &0.017  &0.80  &1.98  &8128  &4.094  &2.090  &17.13  &2.306  &0.696  &0.62  &0.097\\
\hline
\end{tabular}
\end{table*}
\end{document}